\documentclass[10pt,twocolumn,letterpaper]{article}

\usepackage[pagenumbers]{cvpr} % To force page numbers, e.g. for an arXiv version

\usepackage{pifont}
\usepackage{multirow}
\usepackage[ruled,vlined]{algorithm2e}

\definecolor{cvprblue}{rgb}{0.21,0.49,0.74}
\definecolor{darkgreen}{rgb}{0,0.6,0}
\usepackage[pagebackref,breaklinks,colorlinks,allcolors=cvprblue]{hyperref}

\def\paperID{*}
\def\confName{CVPR}
\def\confYear{2025}

\newcommand{\x}{{\mathbf{x}}}
\newcommand{\xhat}{{\hat{\mathbf{x}}}}
\newcommand{\z}{{\mathbf{z}}}
\newcommand{\zhat}{{\hat{\mathbf{z}}}}
\newcommand{\cbf}{{\mathbf{c}}}
\newcommand{\C}{{\mathcal{C}}}
\newcommand{\Enc}{{\mathcal{E}}}
\newcommand{\Ep}{{\boldsymbol{\epsilon}}}
\newcommand{\Dec}{{\mathcal{D}}}
\newcommand{\Loss}{{\mathcal{L}}}

\newcommand{\f}{{\mathbf{f}}}
\newcommand{\Supp}{{\textbf{\textcolor{blue}{Supplementary Material}}}}
\newcommand{\shline}{\specialrule{0.8pt}{0pt}{0pt}}
\newcommand{\best}[1]{\textbf{\textcolor{red}{#1}}}
\newcommand{\sbest}[1]{\underline{\textcolor{blue}{#1}}}
\newcommand{\tbest}[1]{\textit{\textcolor{darkgreen}{#1}}}

\title{Adversarial Diffusion Compression for Real-World Image Super-Resolution}

\author{
\href{https://scholar.google.com/citations?user=aZDNm98AAAAJ}{Bin Chen}$^{1,3,*}$ \qquad
\href{https://github.com/cvsym}{Gehui Li}$^{1,*}$ \qquad
\href{https://scholar.google.com/citations?user=A-U8zE8AAAAJ}{Rongyuan Wu}$^{2,3,*}$ \qquad
\href{https://scholar.google.com/citations?user=q76RnqIAAAAJ}{Xindong Zhang}$^3$ \\
\href{https://aimia-pku.github.io/}{Jie Chen}$^{1,\dagger}$ \qquad
\href{https://jianzhang.tech/}{Jian Zhang}$^{1,\dagger}$ \qquad
\href{http://www4.comp.polyu.edu.hk/~cslzhang/}{Lei Zhang}$^{2,3}$\\
$^1$School of Electronic and Computer Engineering, Peking University\\
$^2$The Hong Kong Polytechnic University \qquad
$^3$OPPO Research Institute\\
{\tt\footnotesize \{chenbin,ligehui921\}@stu.pku.edu.cn} \qquad
{\tt\footnotesize rong-yuan.wu@connect.polyu.hk} \qquad
{\tt\footnotesize zhangxindong1@oppo.com} \\
{\tt\footnotesize \{jiechen2019, zhangjian.sz\}@pku.edu.cn} \qquad
{\tt\footnotesize cslzhang@comp.polyu.edu.hk}}

\begin{document}

\maketitle

\begin{abstract}
Real-world image super-resolution (Real-ISR) aims to reconstruct high-resolution images from low-resolution inputs degraded by complex, unknown processes. While many Stable Diffusion (SD)-based Real-ISR methods have achieved remarkable success, their slow, multi-step inference hinders practical deployment. Recent SD-based one-step networks like OSEDiff and S3Diff alleviate this issue but still incur high computational costs due to their reliance on large pretrained SD models. This paper proposes a novel Real-ISR method, \textbf{AdcSR}, by distilling the one-step diffusion network OSEDiff into a streamlined diffusion-GAN model under our \textbf{A}dversarial \textbf{D}iffusion \textbf{C}ompression (\textbf{ADC}) framework. We meticulously examine the modules of OSEDiff, categorizing them into two types: \textbf{(1) Removable} (VAE encoder, prompt extractor, text encoder, \etc) and \textbf{(2) Prunable} (denoising UNet and VAE decoder). Since direct removal and pruning can degrade the model's generation capability, we pretrain our pruned VAE decoder to restore its ability to decode images and employ adversarial distillation to compensate for performance loss. This ADC-based diffusion-GAN hybrid design effectively reduces complexity by 73\% in inference time, 78\% in computation, and 74\% in parameters, while preserving the model’s generation capability. Experiments manifest that our proposed AdcSR achieves competitive recovery quality on both synthetic and real-world datasets, offering up to 9.3$\times$ speedup over previous one-step diffusion-based methods. Code and models are available at \url{https://github.com/Guaishou74851/AdcSR}.
\end{abstract}

\renewcommand{\thefootnote}{}
\footnote{This work was supported by OPPO Research Fund.}
\footnote{$^*$Equal Contribution. $^\dagger$Corresponding authors.}

\vspace{-5pt}
\section{Introduction}
Image super-resolution (ISR) \cite{dong2014learning,timofte2015a+,zhang2018image,liang2021swinir,liu20242dquant} is a fundamental and long-standing problem in computer vision. It aims to reconstruct the high-resolution (HR) image from a low-resolution (LR) counterpart. One line of ISR research assumes that the LR image $\x_\text{LR}$ is a bicubic-downsampled version of the HR image $\x_\text{HR}$. However, deep ISR networks trained using this assumption often struggle to generalize to real-world scenarios, where degradations are more complex and typically unknown. Another increasingly popular line of ISR research, known as real-world ISR (Real-ISR) \cite{zhang2021designing,wang2021real}, employs random shufflings of degradation operations and high-order degradation processes to synthesize LR-HR training pairs. These approaches have improved the performance of deep ISR networks in real-world scenarios.

\begin{figure}[!t]
\centering
\vspace{-10pt}
\includegraphics[width=\linewidth]{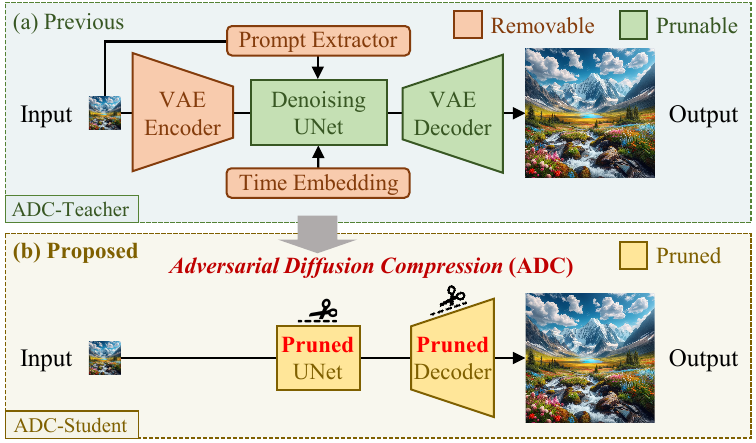}
\vspace{-20pt}
\caption{\textbf{Comparison between our proposed AdcSR and typical one-step diffusion-based Real-ISR methods.} \textbf{(a)} The state-of-the-art one-step diffusion network OSEDiff \cite{wu2024one} employs complete SD \cite{stabilityai} models for Real-ISR, suffering from high computational costs. \textbf{(b)} We distill OSEDiff (ADC-teacher) into a smaller diffusion-GAN hybrid model, \textbf{AdcSR} (ADC-student), under the proposed ADC framework, achieving significantly improved efficiency while maintaining competitive recovery performance.}
\label{fig:teaser}
\vspace{-10pt}
\end{figure}

In the field of ISR and Real-ISR, generative adversarial networks (GANs) \cite{goodfellow2014generative,yang2021gan,wang2021towards,cai2021toward,liang2022details,he2022gcfsr,mou2024empowering,zhang2024pairwise} like SRGAN \cite{ledig2017photo}, BSRGAN \cite{zhang2021designing}, and Real-ESRGAN \cite{wang2021real} have shown greater effectiveness than non-generative models \cite{tong2017image,zhang2018residual,he2021interactive,chen2022content,chen2023activating,chen2024practical,zhang2024progressive,li2024d,li2024osmamba,chen2024self,chen2025invertible} in producing realistic details. In addition to GANs, diffusion \cite{kim2024pagoda,song2020score,dhariwal2021diffusion,chen2024binarized}-based methods such as SR3 \cite{saharia2022image}, StableSR \cite{wang2024exploiting}, and SeeSR \cite{wu2024seesr} have enhanced the quality of super-resolved images by training powerful diffusion networks \cite{li2022srdiff,yue2024resshift,yue2024efficient,luo2023image,delbracio2023inversion,tang2024seeclear} and leveraging pretrained text-to-image (T2I) diffusion models \cite{wang2024exploiting,yang2023pixel,lin2023diffbir,sun2023improving,yu2024scaling,qu2024xpsr,fanadadiffsr} such as Stable Diffusion (SD) \cite{rombach2022high,stabilityai,podell2023sdxl,sd21base}. However, these GANs and diffusion-based Real-ISR approaches suffer from limited recovery quality or slow inference with tens to hundreds of sampling steps.

Recently, efforts \cite{noroozi2024you,he2024one,li2024distillation,kim2024tddsr,xie2024addsr} have been made to improve the inference speed of diffusion models for Real-ISR. For instance, SinSR \cite{wang2024sinsr} distills the 15-step ResShift \cite{yue2024resshift} into a one-step student ISR model. However, it does not utilize large pretrained T2I models and tends to produce oversmoothed results \cite{wu2024one,cui2024taming,zhang2024degradation}. Building on pretrained SD models, OSEDiff \cite{wu2024one} applies variational score distillation (VSD) \cite{wang2024prolificdreamer} to ensure the realism of super-resolution images with a one-step diffusion sampling. S3Diff \cite{zhang2024degradation} designs a degradation-guided Low-Rank Adaptation (LoRA) \cite{hu2021lora} module and an online negative sample generation strategy to improve the perceptual quality of images. Nevertheless, the complexity of these approaches in terms of parameter number and inference time can still be too high for real deployments, especially on resource-limited edge devices.

\begin{figure}[!t]
\centering
\vspace{-10pt}
\includegraphics[width=\linewidth]{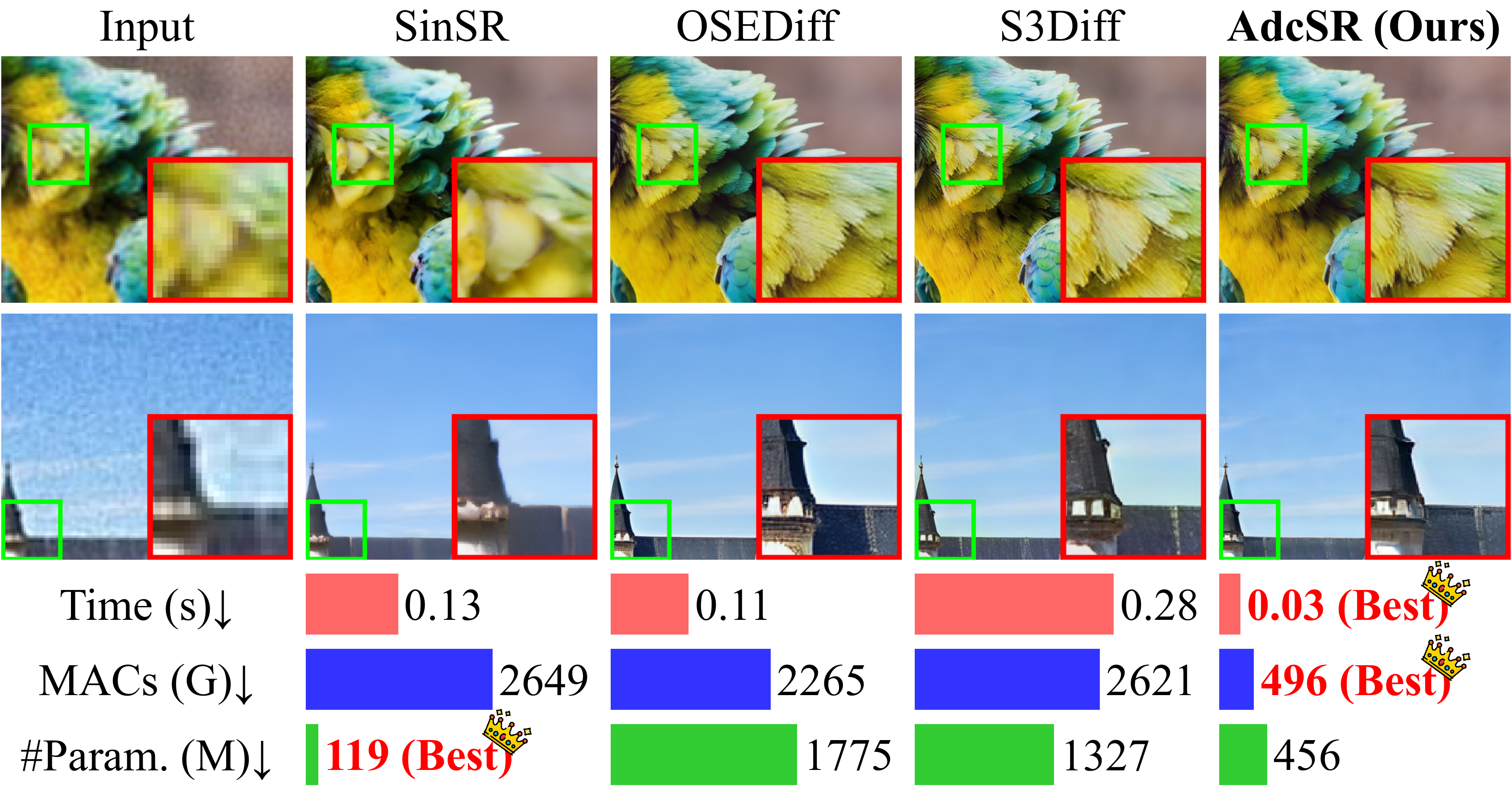}
\vspace{-20pt}
\caption{\textbf{Comparison of our proposed AdcSR with other existing one-step diffusion-based Real-ISR methods} \cite{wang2024sinsr,wu2024one,zhang2024degradation} in terms of visual quality of super-resolution images (top) and model efficiency (bottom). The proposed AdcSR model shows competitive performance in recovering photo-realistic details, while providing the highest inference speed on an NVIDIA A100 GPU, the lowest computational cost, and the second-fewest parameters.}
\label{fig:comp_one_step}
\vspace{-5pt}
\end{figure}

To reduce complexity while maintaining recovery quality, in this paper, we propose a novel diffusion-based Real-ISR model \textbf{AdcSR}, which is obtained by applying our proposed adversarial diffusion compression (ADC) framework to OSEDiff. Our main idea is based on the hypothesis that, given LR input $\x_\text{LR}$ containing abundant information about the target HR image $\x_\text{HR}$, a structurally compressed version of SD-based one-step diffusion networks like OSEDiff \cite{wu2024one} has a sufficient capacity to learn an effective Real-ISR mapping. As illustrated in Fig.~\ref{fig:teaser}, we remove the variational autoencoder (VAE) encoder, prompt extractor, text encoder, cross-attention (CA), and time embedding layers in the SD UNet which we find less important than other modules like self-attention (SA) layers to develop the architecture of AdcSR. Then, we compress the remaining denoising UNet and VAE decoder using channel pruning for improved efficiency. To preserve the model's generative recovery ability while ensuring training efficiency, inspired by the success of diffusion GANs \cite{xiao2021tackling,wang2022diffusion,sauer2023adversarial,kim2023consistency,xu2024ufogen,lin2024sdxl,kang2024distilling,sauer2024fast,luo2024you}, we pretrain our pruned VAE decoder and introduce adversarial distillation in the feature space of VAE decoder. This enables AdcSR to utilize the information from pretrained SD and OSEDiff models, as well as the ground truth (GT) images. By doing so, we significantly reduce the complexity of OSEDiff while maintaining competitive recovery quality, as shown in Fig.~\ref{fig:comp_one_step}. In summary, our contributions are:

\vspace{3pt}
\noindent \ding{113} (1) We introduce ADC, a novel framework that combines structural compression (module removal and pruning) with adversarial distillation (knowledge distillation with adversarial loss) to streamline SD-based one-step Real-ISR models into smaller diffusion-GAN hybrid networks.

\vspace{3pt}
\noindent \ding{113} (2) We design a structural compression strategy in ADC: firstly, removing unnecessary modules (VAE encoder, text, and time modules), and then pruning the remaining compressible modules (denoising UNet and VAE decoder).

\vspace{3pt}
\noindent \ding{113} (3) We develop a two-stage training scheme in our ADC: firstly, pretraining a channel-pruned VAE decoder, and then distilling one-step teacher into our model with an adversarial loss in the feature space of pretrained VAE decoder.

\vspace{3pt}
\noindent \ding{113} (4) By applying ADC to a state-of-the-art SD-based one-step network \cite{wu2024one}, we propose AdcSR model, a structurally compressed diffusion GAN that effectively achieves a 3.7$\times$ inference acceleration and a 74\% reduction in parameters.

\vspace{3pt}
\noindent \ding{113} (5) Experiments exhibit the competitive Real-ISR performance of our AdcSR model and its appealing efficiency.

\section{Related Work}
\label{sec:related_work}
\noindent \textbf{Real-ISR based on LR-HR Pair Synthesis.} To make ISR networks applicable to real scenarios, BSRGAN \cite{zhang2021designing} and Real-ESRGAN \cite{wang2021real} pioneer the use of shuffled and high-order degradations to synthesize LR-HR pairs for training Real-ISR GANs. They inspire a lot of works \cite{chen2022real,liang2022details,xie2023desra,zhang2024real} that develop new degradation prediction mechanisms \cite{liang2022efficient,mou2024empowering} and network structures \cite{liang2021swinir,chen2023activating}. However, these approaches often suffer from artifacts and oversmoothing.

The success of diffusion models in high-quality generation has prompted researchers to explore leveraging powerful diffusion priors like SD \cite{rombach2022high,stabilityai} for Real-ISR. Most SD-based methods \cite{lin2023diffbir,sun2023improving,yu2024scaling} train adapter modules \cite{zhang2023adding,mou2024t2i} that use the LR image as control signal to guide the super-resolution processes. For example, StableSR \cite{wang2024exploiting} finetunes a time-aware encoder and introduces a controllable feature warping module to balance quality and fidelity. PASD \cite{yang2023pixel} extracts both low-level and high-level features from the LR image and inputs them into the pretrained SD model with a pixel-aware CA module. SeeSR \cite{wu2024seesr} enhances model's semantic awareness by using degradation-robust tag-style text prompts and soft prompts to guide diffusion sampling. In addition to these, ResShift \cite{yue2024resshift} introduces a new residual shifting-based diffusion model to improve the efficiency of the transition from $\x_\text{LR}$ to $\x_\text{HR}$. However, these approaches require tens to hundreds of iterative steps for diffusion sampling, which increases inference latency and limits their application in real deployments where fast inference is critical.

\vspace{3pt}
\noindent \textbf{Diffusion Distillation for One-Step Inference.} To accelerate the generation process of diffusion models, numerous techniques \cite{luhman2021knowledge,salimans2022progressive,meng2023distillation,heek2024multistep,yan2024perflow,luhman2021knowledge,ren2024hyper,xu2024accelerating,zheng2024trajectory,gu2023boot} have been proposed to distill a multi-step diffusion sampling process into a student model with fewer steps. Recent methods \cite{berthelot2023tract,zhu2024slimflow} further reduce the required number of steps to just one. For instance, InstaFlow \cite{liu2023instaflow} distills an ordinary differential equation (ODE) sampling trajectory into a one-step network. Consistency models \cite{song2023consistency,luo2023latent} learn to output consistent results at any timestep. Subsequent works like CTM \cite{kim2023consistency}, SDXL-Lightning \cite{lin2024sdxl}, UFOGen \cite{xu2024ufogen}, LADD \cite{sauer2023adversarial,sauer2024fast}, DMD2 \cite{yin2024one}, and Diffusion2GAN \cite{kang2024distilling} leverage adversarial distillation to improve the quality of generated images using pretrained networks as discriminators. For Real-ISR, SinSR \cite{wang2024sinsr} shortens ResShift \cite{yue2024resshift} via bidirectional distillations. OSEDiff \cite{wu2024one} introduces VSD \cite{wang2024prolificdreamer} approach in latent space to enhance the realism of super-resolved images. Building upon the distilled SD-Turbo \cite{sauer2023adversarial} models, S3Diff \cite{zhang2024degradation} designs a degradation-guided LoRA module and an online negative prompting strategy for improved ISR quality. However, the complexity of existing SD-based one-step diffusion networks remains too high for real deployment on mobile and edge devices due to their large-scale parameters and heavy computation. To mitigate this problem, we structurally compress and distill OSEDiff into a smaller diffusion GAN, enhancing efficiency while maintaining performance.

\vspace{3pt}  
\noindent \textbf{Structural Compression for Latent Diffusion Models.} To achieve photo-realistic image generation, large-scale latent diffusion models \cite{rombach2022high,stabilityai,podell2023sdxl} are widely employed due to their powerful generative priors. However, the deployment of these models is hindered by their high computation costs. To address this issue, a lot of works \cite{fang2023structural,castells2024edgefusion,zhu2024slimflow,zhang2024laptop,zhao2023mobilediffusion} have explored compression techniques for efficiency. For example, BK-SDM \cite{kim2023bk} applies block removal for SD models. SnapFusion \cite{li2024snapfusion} designs block-removed UNet and efficient VAE decoder with an improved distillation approach, achieving 8-step T2I inferences. To our knowledge, no existing compression techniques are specifically designed for diffusion-based Real-ISR. In this work, we propose a novel method based on introduced adversarial diffusion compression (ADC). Moving beyond previous one-step approaches \cite{wang2024sinsr,wu2024one,cui2024taming,zhang2024degradation,li2024distillation}, we demonstrate that, given LR image as a starting point of super-resolution, the latent encoding, prompt extraction, text-conditioned denoising, and decoding can be compressed into an optimized diffusion GAN.

\begin{figure*}[!t]
\vspace{-5pt}
\centering
\includegraphics[width=1.02\textwidth]{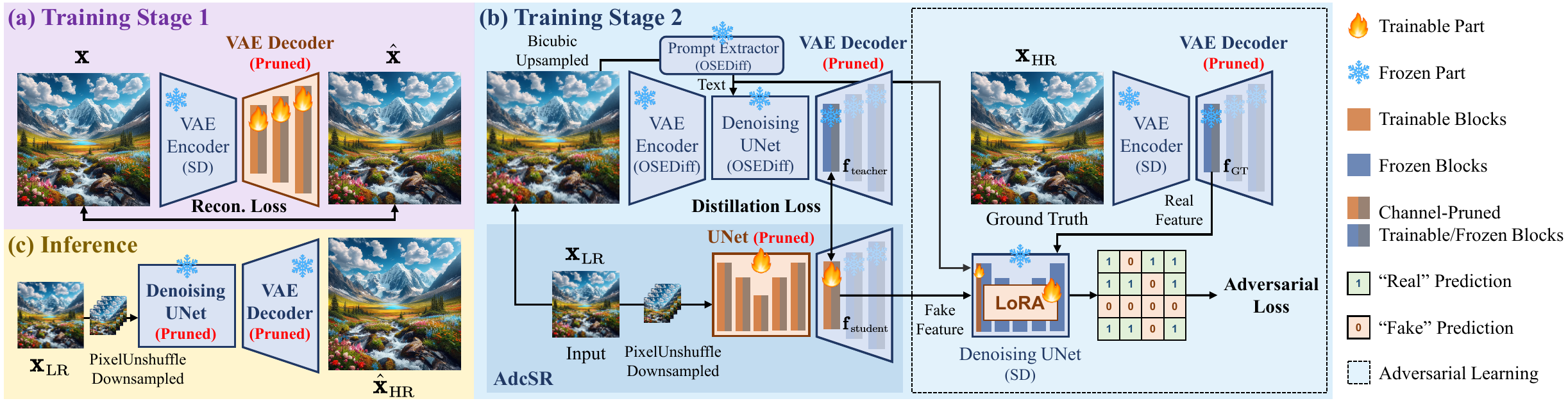}
\vspace{-20pt}
\caption{\textbf{Illustration of the training and inference processes of AdcSR, an instantiation of our ADC framework applied to OSEDiff.} \textbf{(a)} In Stage 1, we pretrain a pruned VAE decoder that shares the latent space with SD and OSEDiff. \textbf{(b)} In Stage 2, we distill the knowledge from OSEDiff (\textcolor{blue}{\textbf{ADC-teacher}}) into AdcSR (\textcolor{blue}{\textbf{ADC-student}}) by aligning features in the pretrained decoder. An adversarial loss encourages the student to generate features that can fool a LoRA-finetuned SD UNet (\textcolor{blue}{\textbf{ADC-discriminator}}), utilizing the corresponding real features of GT images. Since all supervisions perform in the feature space, there is no need to decode images as in previous approaches \cite{wu2024one,zhang2024degradation}. \textbf{(c)} During inference, the LR image is directly fed into our trained compressed UNet and VAE decoder to obtain the super-resolution result.}
\label{fig:method}
\vspace{-5pt}
\end{figure*}

\section{Method}
\subsection{Preliminary}
\noindent \textbf{OSEDiff, and Its Limitations.} OSEDiff \cite{wu2024one} is a typical state-of-the-art one-step diffusion-based Real-ISR method that employs a LoRA-finetuned SD VAE encoder $\Enc_\text{OSEDiff}$, a LoRA-finetuned SD UNet $\Ep_\text{OSEDiff}$, a pretrained SD VAE decoder $\Dec_\text{SD}$, and a pretrained prompt extractor $\C$ \cite{wu2024seesr} to perform super-resolution through the following process:
\begin{align}
& \z_\text{LR} = \Enc_\text{OSEDiff}(\x_\text{LR}), \quad \cbf = \C(\x_\text{LR}),  \label{eq:osediff_encode_extract} \\  
& \zhat_\text{HR} = \left[\z_\text{LR} - \sqrt{1-\bar{\alpha}_{T}} \Ep_\text{OSEDiff}(\z_\text{LR};T,\cbf)\right]/{\sqrt{\bar{\alpha}_{T}}}, \label{eq:osediff_denoising} \\  
& \xhat_\text{HR} = \Dec_\text{SD}(\zhat_\text{HR}). \label{eq:osediff_decode}
\end{align}
In Eq.~(\ref{eq:osediff_encode_extract}), the LR image $\x_\text{LR}$ is encoded into the VAE latent space, and text prompts $\cbf$ are extracted from $\x_\text{LR}$ in parallel. In Eq.~(\ref{eq:osediff_denoising}), one-step diffusion denoising is executed using the noise schedule $\{\bar{\alpha}_t\}$ \cite{ho2020denoising} at the $T$-th timestep. Finally, in Eq.~(\ref{eq:osediff_decode}), the denoised latent code is decoded back into image space to obtain the super-resolution image $\xhat_\text{HR}$. However, OSEDiff has a total parameter number of 1775M and an inference latency of 0.11s on an NVIDIA A100 GPU for a $512 \times 512$ target HR image, which can still be too expensive for real deployment environments where both computational and storage resources are limited. Similar challenges persist in other one-step diffusion-based methods utilizing large-scale pretrained SD models \cite{cui2024taming,zhang2024degradation,noroozi2024you,li2024distillation,xie2024addsr}.

\subsection{Structural Compression Strategy}
To improve the efficiency of SD-based Real-ISR methods, we propose an \textbf{A}dversarial \textbf{D}iffusion \textbf{C}ompression (\textbf{ADC}) framework. Its key insight is that ISR differs from T2I tasks, which rely solely on text inputs for generation, while the LR image in Real-ISR provides rich information about the target HR image. Thus, unlike previous SD-based one-step approaches that employ complete SD model structures, we hypothesize that competitive Real-ISR performance does not require these full architectures, which have been validated to possess sufficient capacity for one-step T2I and Real-ISR (see Sec.~\ref{sec:related_work}). Taking OSEDiff \cite{wu2024one} as example in this work, we propose that the modules used in Eqs.~(\ref{eq:osediff_encode_extract})-(\ref{eq:osediff_decode}) contain redundancy and can be removed or pruned for efficiency. To be specific, as shown in Fig.~\ref{fig:teaser}, we categorize the modules into two types: \textbf{(1) Removable} (VAE encoder, prompt extractor, text encoder, CA layers, and time embeddings) and \textbf{(2) Prunable} (denoising UNet and VAE decoder). Based on this categorization, in ADC, we design a structural compression strategy that includes two modifications for SD-based one-step methods: \textbf{(1) Removal} of unnecessary modules, and \textbf{(2) Pruning} of remaining compressible modules. In the following, we detail and justify these modifications.

\subsubsection{Removal of Unnecessary Modules}
\noindent \textbf{Eliminating VAE Encoder.} In previous SD-based one-step Real-ISR approaches, the VAE encoder maps $\mathbf{x}_\text{LR}$ to a latent code $\z_\text{LR}$, as shown in Eq.~(\ref{eq:osediff_encode_extract}). This process involves multiple downsampling operations, which can lead to the loss of information important for Real-ISR. To preserve the complete information of the LR input without loss, we eliminate the VAE encoder entirely. Instead, we apply a PixelUnshuffle \cite{shi2016real} operation to $\mathbf{x}_\text{LR}$, rearranging its spatial pixels into channel dimension while maintaining the same spatial size as $\z_\text{LR}$. Correspondingly, the first convolution of UNet is adjusted to match the increased channel number, and the output of PixelUnshuffle is then directly input into the UNet.

\vspace{3pt}
\noindent \textbf{Removing Text and Time Modules.} In models like OSEDiff, a prompt extractor generates textual prompts from $\mathbf{x}_\text{LR}$, which are then used in text encoder and CA layers within the denoising UNet, as shown in Eqs. (\ref{eq:osediff_encode_extract}) and (\ref{eq:osediff_denoising}). Additionally, time embeddings are included to condition the UNet on different timesteps. While the text prompts are generally important for guiding T2I synthesis, in the specific context of Real-ISR, we have empirically observed that they contribute less significantly to enhance quality, compared to the other remaining modules. Furthermore, since OSEDiff performs only one-step diffusion sampling, time embeddings are unnecessary, as there is no need to differentiate between timesteps. Therefore, we remove the prompt extractor, text encoder, CA layers, and time embeddings from the UNet, retaining only its SA, linear, and convolutional layers.

\begin{table*}[!t]
\vspace{-5pt}
\setlength{\tabcolsep}{5pt}
\centering
\caption{\textbf{Quantitative comparison of different methods on DRealSR.} Efficiency metrics are tested on an NVIDIA A100 GPU. Throughout this paper, the best, second-best, and third-best results are highlighted in \best{bold red}, \sbest{underlined blue}, and \tbest{italic green}, respectively.}
\vspace{-5pt}
\resizebox{\linewidth}{!}{
\begin{tabular}{l|ccccccc|>{\columncolor[HTML]{FFEEED}}c>{\columncolor[HTML]{FFEEED}}c>{\columncolor[HTML]{FFEEED}}c>{\columncolor[HTML]{FFEEED}}c}
\shline
Method & PSNR↑ & SSIM↑ & LPIPS↓ & DISTS↓ & NIQE↓ & MUSIQ↑ & CLIPIQA↑ & \#Steps↓ & Time (s)↓ & MACs (G)↓ & \#Param. (M)↓ \\
\hline \hline
StableSR \cite{wang2024exploiting} & 28.03 & 0.7536 & 0.3284 & 0.2269 & 6.52 & 58.51 & 0.6356 & 200 & 11.50 & 79940 & 1410 \\
DiffBIR \cite{lin2023diffbir} & 26.71 & 0.6571 & 0.4557 & 0.2748 & \tbest{6.31} & 61.07 & 0.6395 & 50 & 2.72 & 24234 & 1717 \\
SeeSR \cite{wu2024seesr} & \tbest{28.17} & \tbest{0.7691} & 0.3189 & 0.2315 & 6.40 & \sbest{64.93} & 0.6804 & 50 & 4.30 & 65857 & 2524 \\
PASD \cite{yang2023pixel} & 27.36 & 0.7073 & 0.3760 & 0.2531 & \best{5.55} & \tbest{64.87} & 0.6808 & \tbest{20} & 2.80 & 29125 & 1900 \\
ResShift \cite{yue2024resshift} & \best{28.46} & 0.7673 & 0.4006 & 0.2656 & 8.12 & 50.60 & 0.5342 & \sbest{15} & 0.71 & 5491 & \best{119} \\
\hline
SinSR \cite{wang2024sinsr} & \sbest{28.36} & 0.7515 & 0.3665 & 0.2485 & 6.99 & 55.33 & 0.6383 & \best{1} & \tbest{0.13} & 2649 & \best{119} \\
OSEDiff \cite{wu2024one} & 27.92 & \best{0.7835} & \best{0.2968} & \sbest{0.2165} & 6.49 & 64.65 & \tbest{0.6963} & \best{1} & \sbest{0.11} & \sbest{2265} & 1775 \\
S3Diff \cite{zhang2024degradation} & 27.39 & 0.7469 & \tbest{0.3129} & \best{0.2108} & \sbest{6.17} & 64.16 & \best{0.7156} & \best{1} & 0.28 & \tbest{2621} & \tbest{1327} \\
\textbf{AdcSR (Ours)} & 28.10 & \sbest{0.7726} & \sbest{0.3046} & \tbest{0.2200} & 6.45 & \best{66.26} & \sbest{0.7049} & \best{1} & \best{0.03} & \best{496} & \sbest{456} \\
\shline
\end{tabular}}
\label{tab:comp_diff_quantitative}
\vspace{-5pt}
\end{table*}

\subsubsection{Pruning of Remaining Modules}
\noindent \textbf{Optimizing UNet-VAE Decoder Connection.} Before decoding the output image, traditional SD-based methods like OSEDiff map the high-capacity feature (often hundreds of channels) in UNet to a 4-channel latent code $\zhat_\text{HR}$. This dimensionality reduction can potentially result in a loss of feature information and constrain the model's representation ability. To mitigate this and fully leverage the rich feature representations learned by the UNet, we enhance the information flow between UNet and VAE decoder. Specifically, we remove the output layer of UNet and the input layer of VAE decoder, which reduce and then increase the feature channels. Instead, we introduce a convolution layer that directly connects the high-dimensional feature in UNet to the first blocks of the VAE decoder, improving the model's recovery quality while reducing its overall inference latency.

\vspace{3pt}
\noindent \textbf{Pruning Feature Channels.} We hypothesize that the current one-step model, compressed by the above three operations, still contains redundancy and has sufficient capacity to learn an effective Real-ISR mapping with even fewer parameters. Although previous works \cite{song2024sdxs,kim2023bk,castells2024edgefusion,zhang2024laptop,ma2024deepcache} compress SD-based models by removing network blocks or layers, we find that this can noticeably degrade the performance of one-step diffusion networks, where the depth of UNet and VAE decoder is already relatively shallow. Further decreasing the depth may impair the ability of model to extract hierarchical features and learn complex transformations for high-quality Real-ISR. To avoid this issue and strike a balance between recovery quality and efficiency, we opt for channel pruning. Concretely, we retain 75\% of the feature channels in the UNet and 50\% channels in the VAE decoder. This reduces the model's complexity while alleviating performance loss by preserving network depth.

\vspace{3pt}
The resulting structurally compressed model, which we name \textbf{AdcSR}, incorporates the proposed two modifications in ADC and consists of three modules: \textbf{(1)} a PixelUnshuffle layer that prepares the LR input image $\x_\text{LR}$ for processing by rearranging its pixels without information loss; \textbf{(2)} a channel-pruned SD UNet without text encoder, CA layers, and time embeddings, processing the rearranged LR image while keeping the original depth; and \textbf{(3)} a channel-pruned VAE decoder which receives the high-dimensional features from UNet and generates the super-resolution image $\xhat_\text{HR}$.

\begin{figure}[!t]
\vspace{-5pt}
\centering
\includegraphics[width=0.97\linewidth]{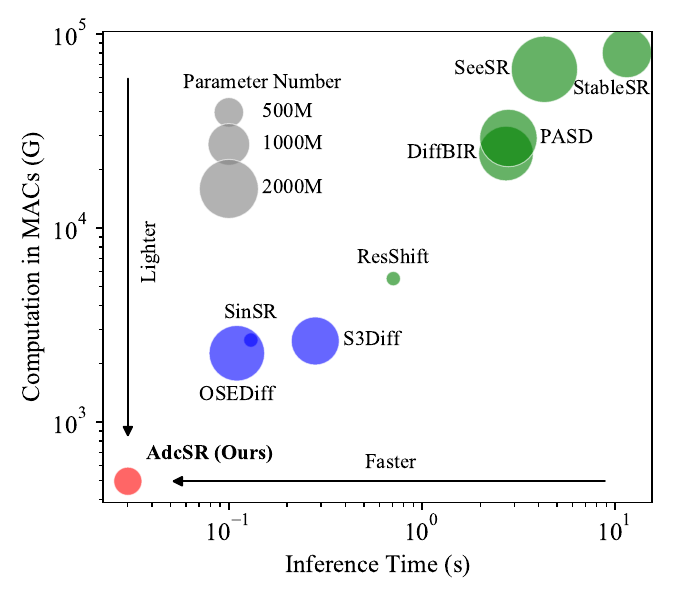}
\vspace{-12.5pt}
\caption{\textbf{Efficiency comparison} using a bubble plot, showing the inference time, computation, and parameter number (see Tab.~\ref{tab:comp_diff_quantitative}) for super-resolving a $128\times 128$ LR image on an NVIDIA A100 GPU. AdcSR achieves the fastest inference, lightest computation, and second-fewest parameters. Bubble colors represent approach types: green for multi-step, blue for one-step, and red for AdcSR.}
\label{fig:comp_complexity}
\vspace{-5pt}
\end{figure}

\subsection{Training Scheme}
Direct removal and pruning can degrade the model's generative capabilities due to reduced capacity and altered network structure. To mitigate this, as Fig.~\ref{fig:method} shows, our ADC uses a two-stage training scheme: (1) pretraining VAE decoder, and (2) adversarial distillation to compensate for potential performance loss and ensure high-quality Real-ISR.

\vspace{3pt}
\noindent \textbf{Stage 1: Pretraining Channel-Pruned VAE Decoder.} In the first stage, we pretrain a pruned VAE decoder \cite{van2017neural,esser2021taming} to restore its ability to decode images. As shown in Fig.~\ref{fig:method} (a), we freeze the parameters of the pretrained SD VAE encoder and train only the VAE decoder from scratch. Given an input image $\x$, the encoder produces latent codes, which are then decoded back into an image $\xhat$ by the decoder. To train the decoder, following \cite{rombach2022high}, we adopt a reconstruction loss consisting of a pixel-level $L_1$ loss $\lVert \xhat - \x \rVert_{1}$, an LPIPS loss \cite{zhang2018unreasonable}, and a patch-based adversarial loss \cite{isola2017image,dosovitskiy2016generating,esser2021taming} to encourage the reconstructed $\xhat$ to be visually similar to $\x$.

\vspace{3pt}
\noindent \textbf{Stage 2: Knowledge Distillation with Adversarial Loss.} In the second stage, we distill the knowledge from the pretrained OSEDiff (teacher) into our compressed AdcSR (student). Specifically, as illustrated in Fig.~\ref{fig:method} (b), we connect the pruned UNet and all the first blocks of pruned decoder at the level with the smallest spatial size, and jointly finetune them. The student is initialized using the pretrained SD and VAE decoder from Stage 1. Distillation is performed in the feature space by aligning the student's features $\f_{\text{student}}$ with the teacher's corresponding features $\f_{\text{teacher}}$ using an $L_1$ loss:
\begin{equation}
\Loss_{\text{distill}} = \lVert \mathbf{f}_{\text{student}} - \mathbf{f}_{\text{teacher}} \rVert_{1}.
\end{equation}
Here, $\f_{\text{teacher}}$ is obtained by passing the LR image through the teacher's VAE encoder, prompt extractor, UNet, and all first blocks of the pretrained decoder, while $\f_{\text{student}}$ is produced from the student's pruned UNet and all first blocks of the pruned decoder. This distillation in feature space is both effective and efficient without the need to decode images.

To further enhance the visual quality of super-resolution outputs, we introduce an adversarial loss on $\f_{\text{student}}$, encouraging it to follow the same distribution as the corresponding features of GT images. Specifically, we obtain the real features $\f_{\text{GT}}$ by encoding $\x_{\text{HR}}$ using SD encoder and processing them with the first blocks of pruned decoder at the smallest spatial size. We reuse a pretrained SD UNet as the discriminator, where the first convolution layer is adjusted to match the channel number of $\f_{\text{student}}$ and $\f_{\text{GT}}$. In addition, we integrate LoRA modules, ensuring that only the LoRA and the first convolution layer remain trainable, while all other parameters are frozen to efficiently finetune the pretrained SD UNet. The discriminator is conditioned on the text prompts $\cbf$ extracted by the teacher, with timestep fixed at $T$. Following \cite{yin2024improved}, we employ the non-saturating adversarial loss:
\begin{equation}
\Loss_\text{adv} = \text{Softplus} \left( -\text{Discriminator}(\f_\text{student}) \right),
\end{equation}
which provides fine-grained feedback as the discriminator's output shares the same spatial dimension as input features. The total training loss is defined as $\Loss =\Loss_\text{distill} + \lambda_\text{adv}\Loss_\text{adv}$.

\begin{figure*}[!t]
\setlength{\tabcolsep}{0.5pt}
\centering
\footnotesize
\resizebox{\linewidth}{!}{
\begin{tabular}{cccccc}
Input & DiffBIR \cite{lin2023diffbir} & SeeSR \cite{wu2024seesr} & SinSR \cite{wang2024sinsr} & OSEDiff \cite{wu2024one} & \textbf{AdcSR \selectfont (Ours)}\\
\includegraphics[width=0.16\textwidth]{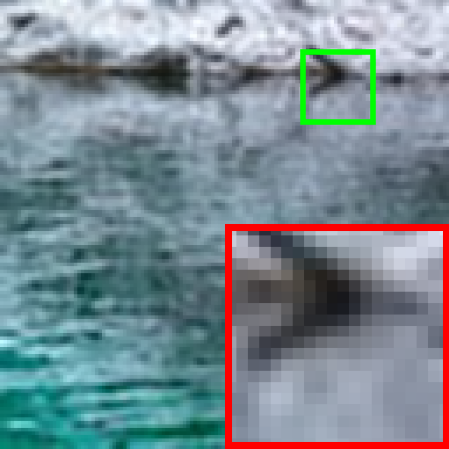}
&\includegraphics[width=0.16\textwidth]{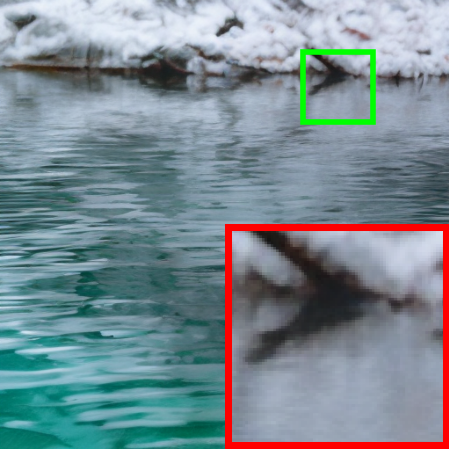}
&\includegraphics[width=0.16\textwidth]{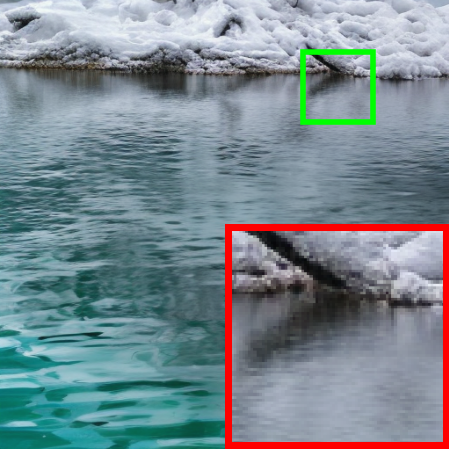}
&\includegraphics[width=0.16\textwidth]{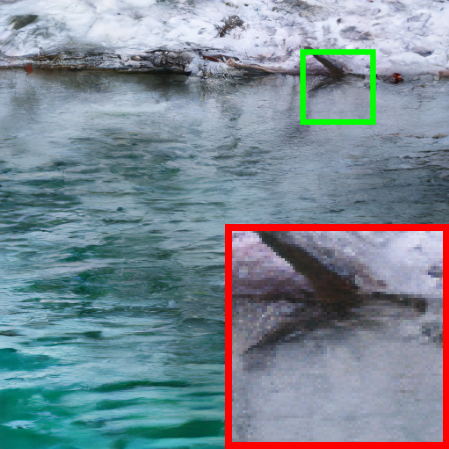}
&\includegraphics[width=0.16\textwidth]{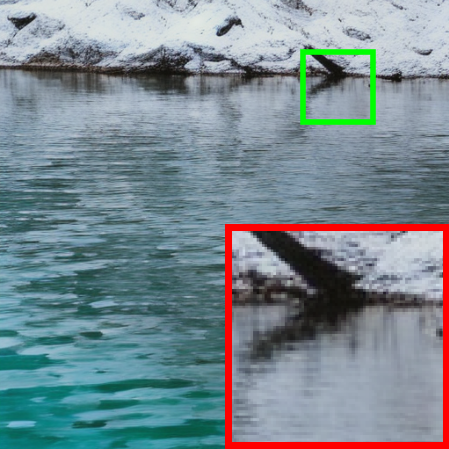}
&\includegraphics[width=0.16\textwidth]{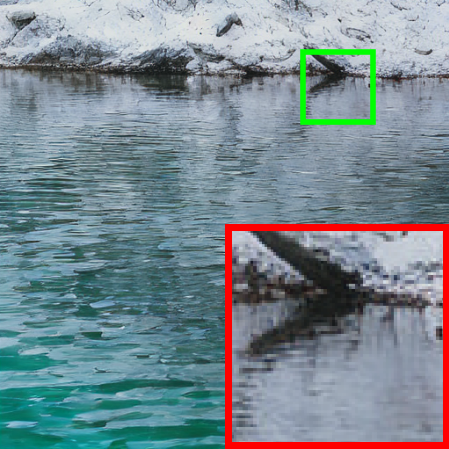}\\
Input & StableSR \cite{wang2024exploiting} & PASD \cite{yang2023pixel} & ResShift \cite{yue2024resshift} & S3Diff \cite{zhang2024degradation} & \textbf{AdcSR (Ours)}\\
\includegraphics[width=0.16\textwidth]{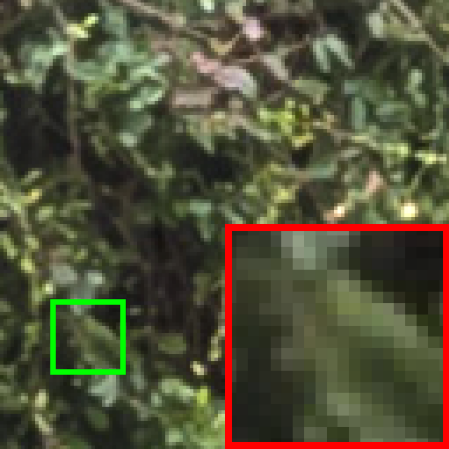}
&\includegraphics[width=0.16\textwidth]{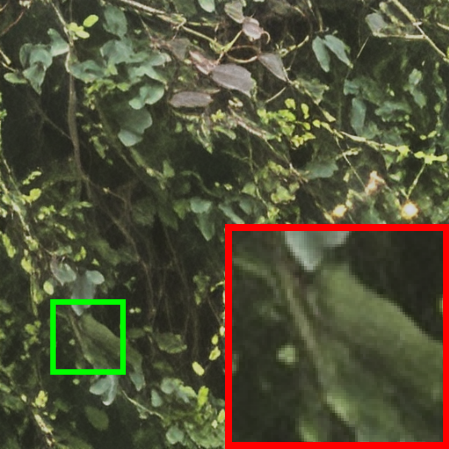}
&\includegraphics[width=0.16\textwidth]{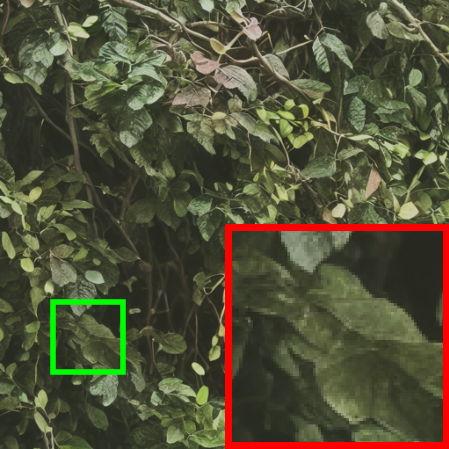}
&\includegraphics[width=0.16\textwidth]{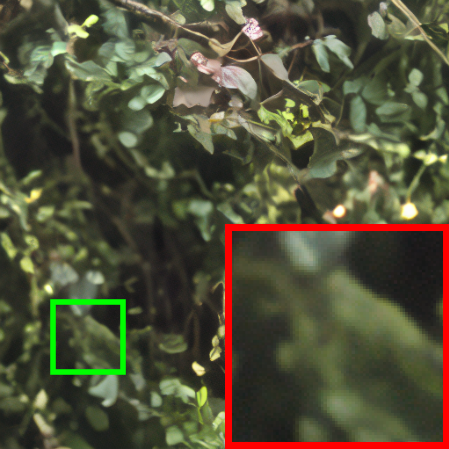}
&\includegraphics[width=0.16\textwidth]{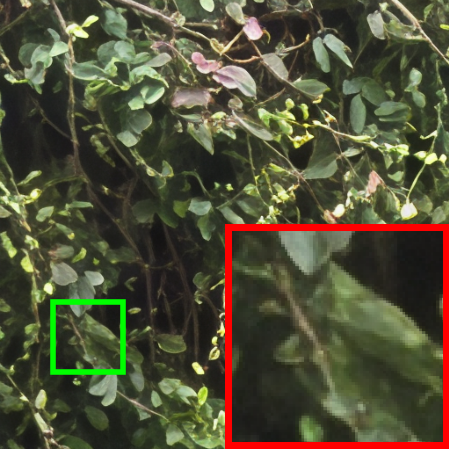}
&\includegraphics[width=0.16\textwidth]{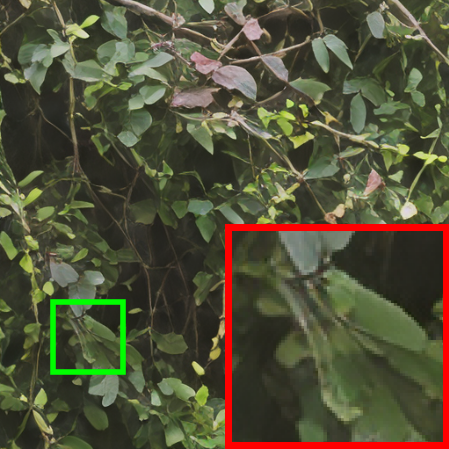}\\
\end{tabular}}
\vspace{-10pt}
\caption{\textbf{Qualitative comparison} on images named ``0835\_pch\_00035" from DIV2K-Val (top) and ``Nikon\_045" from RealSR (bottom).}
\label{fig:comp_diff_qualitative}
\vspace{-5pt}
\end{figure*}

\section{Experiment}
\subsection{Experimental Setting}
\label{subsec:setting}
\noindent \textbf{Implementation Details.} Following \cite{wu2024seesr,wang2024sinsr,wu2024one,zhang2024degradation,sun2023improving,wang2024exploiting,yue2024resshift,lin2023diffbir,yang2023pixel,cui2024taming,li2024distillation,xie2024addsr}, we conduct experiments on the Real-ISR task with scaling factor 4. The sizes of LR and HR images are set to $128\times 128$ and $512\times 512$ by default. We initialize our pruned SD UNet using the pretrained weights of SD2.1-base \cite{sd21base}, reusing only the parameters corresponding to the first 75\% of intermediate feature channels while removing the rest. To match the $64\times 64$ spatial size of latent codes in SD, we set the scaling factor of PixelUnshuffle layer to 2. The convolutional kernels and biases in the first and last UNet layers are repeated in the channel dimension to align with the rearranged LR image and the intermediate features of the first blocks in our pruned SD VAE decoder.

In Stage 1, we employ the code of latent diffusion models \cite{rombach2022high,ldm} to pretrain a 50\% channel-pruned SD VAE decoder from scratch on OpenImage \cite{openimage} for 250K steps, followed by 250K steps on LAION-Face \cite{laionface} and LAION-Aesthetic \cite{laionaesthetics}. The weighting factors of $L_1$ loss and LPIPS loss are both set to 1, while the weighting factor of the patch-based adversarial loss is set to 0 for the first 50K steps and 1 for the remaining steps. The learning rate is fixed at 1.3e-6.

In Stage 2, we jointly finetune the 25\% channel-pruned UNet and all first blocks at the smallest spatial size of the pretrained VAE decoder from Stage 1 on LSDIR \cite{li2023lsdir} with $\lambda_\text{adv}=1$ for 200K steps. The learning rate is initialized at 1e-4 and halved for every 100K steps. The learning rate and LoRA rank for the discriminator are set to 1e-6 and 4, respectively. The high-order degradation pipeline of Real-ESRGAN \cite{wang2021real} is used to synthesize LR-HR pairs. In both two stages, we employ the Adam \cite{kingma2014adam} optimizer and a batch size of 96 for training on 8 NVIDIA A100 (80GB) GPUs.

\vspace{3pt}
\noindent \textbf{Test Datasets.} Following \cite{wu2024seesr,wu2024one,zhang2024degradation}, we test AdcSR and compare it with other methods using the 3K synthesized test images from DIV2K-Val \cite{agustsson2017ntire,wang2024exploiting} and the center-cropped real images from RealSR \cite{cai2019toward} and DRealSR \cite{wei2020component}.

\vspace{3pt}
\noindent \textbf{Compared Methods.} We compare the proposed AdcSR model against eight diffusion-based approaches: StableSR \cite{wang2024exploiting}, DiffBIR \cite{lin2023diffbir}, SeeSR \cite{wu2024seesr}, PASD \cite{yang2023pixel}, ResShift \cite{yue2024resshift}, SinSR \cite{wang2024sinsr}, OSEDiff \cite{wu2024one}, and S3Diff \cite{zhang2024degradation}.

\vspace{3pt}  
\noindent \textbf{Evaluation Metrics.} We adopt both full- and no-reference metrics for performance evaluation. For reference-based fidelity, we use PSNR and SSIM \cite{wang2004image}, calculated on the Y channel in the YCrCb space. For reference-based perceptual quality, we apply LPIPS \cite{zhang2018unreasonable} and DISTS \cite{ding2020image}. FID \cite{heusel2017gans} is also employed to measure the distance between the distributions of GT and super-resolution images. In addition, we utilize no-reference metrics including NIQE \cite{zhang2015feature}, MUSIQ \cite{ke2021musiq}, MANIQA \cite{yang2022maniqa}, and CLIPIQA \cite{wang2023exploring}.

\subsection{Comparison with State-of-the-Arts}  
\noindent \textbf{\fontsize{9.25pt}{12pt}\selectfont Recovery Quality Comparison.} The first 8 columns of Tab.~\ref{tab:comp_diff_quantitative} manifest that our AdcSR achieves promising results across multiple metrics. Firstly, it ranks in top 3 for full-reference quality metrics SSIM, LPIPS, and DISTS, surpassing most other approaches. Secondly, it attains competitive results in PSNR and no-reference metrics NIQE, MUSIQ, and CLIPIQA, performing on par with many state-of-the-art methods. Thirdly, compared to the previous one-step diffusion-based models SinSR and particularly its teacher OSEDiff, AdcSR yields superiority in most of the perceptual quality metrics, and remains competitive with S3Diff across various cases.

Figs.~\ref{fig:comp_one_step} (top) and \ref{fig:comp_diff_qualitative} exhibit the competitive performance of AdcSR in recovering sharp and photo-realistic images. We observe that StableSR, DiffBIR, SeeSR, and PASD can bring unnatural artifacts and blurriness at the intersection of the rocky landscape and the water, along with noise and distortions in the regions of leaves. ResShift and SinSR suffer from noticeable blurry artifacts. OSEDiff and S3Diff could generate fewer details on the surfaces of rocks and water, introducing an additional slight highlight effect on the cluster of leaves. In comparison, AdcSR effectively reconstructs vivid details and natural textures in the regions of parrot's feathers, building, rocky landscape, still water, and leaves.

\begin{table}[!t]
\vspace{-5pt}
\setlength{\tabcolsep}{2.5pt}
\centering
\caption{\textbf{\fontsize{8.5pt}{12pt}\selectfont Ablation study of eliminating VAE encoder} on DRealSR.}
\vspace{-5pt}
\resizebox{\linewidth}{!}{
\begin{tabular}{l|ccccc}
\shline
Method & PSNR↑ & LPIPS↓ & DISTS↓ & \#Param. (M)↓ & Time (s)↓ \\
\hline \hline
w/o Elimination & 27.97 & 0.3077 & 0.2239 & 490 & 0.05 \\
\textbf{w/ Elim. (Ours)} & \best{28.10} & \best{0.3046} & \best{0.2200} & \best{456} & \best{0.03} \\
\shline
\end{tabular}}
\label{tab:abla_elim_enc}
\vspace{-5pt}
\end{table}

\begin{table}[!t]
\setlength{\tabcolsep}{3pt}
\centering
\caption{\textbf{Ablation study of optimizing the connection between the denoising UNet and the VAE decoder} on DRealSR.}
\vspace{-5pt}
\resizebox{\linewidth}{!}{
\begin{tabular}{l|ccccc}
\shline
Method & FID↓ & MUSIQ↑ & MANIQA↑ & CLIPIQA↑ & FPS↑ \\
\hline \hline
w/o Optimization & 140.09 & 65.18 & 0.5807 & 0.6756 & 34.66 \\
\textbf{w/ Opt. (Ours)} & \best{134.05} & \best{66.26} & \best{0.5927} & \best{0.7049} & \best{34.79} \\
\shline
\end{tabular}}
\label{tab:abla_opt_conn}
\vspace{-5pt}
\end{table}

\begin{figure}[!t]
\setlength{\tabcolsep}{0.5pt}
\centering
\resizebox{\linewidth}{!}{
\scriptsize
\begin{tabular}{ccccc}
Input & w/o Elim. & w/o Opt. & \textbf{Ours} & GT\\
\includegraphics[width=0.08\textwidth]{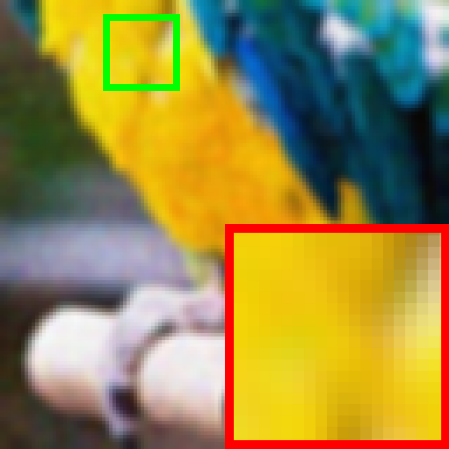}
&\includegraphics[width=0.08\textwidth]{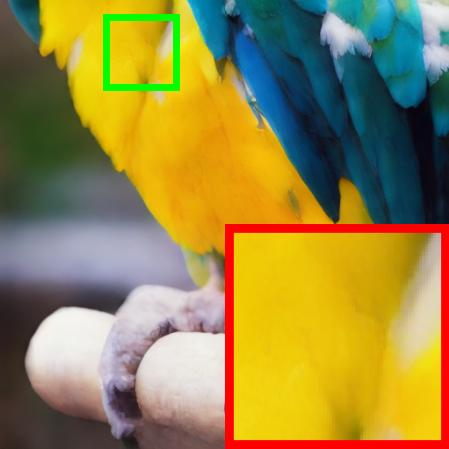}
&\includegraphics[width=0.08\textwidth]{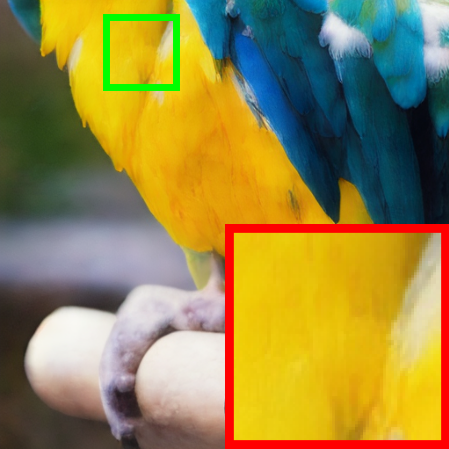}
&\includegraphics[width=0.08\textwidth]{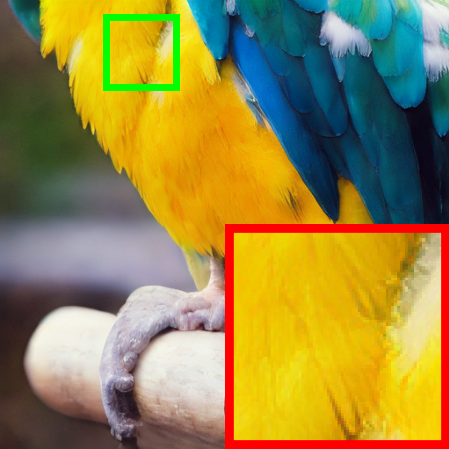}
&\includegraphics[width=0.08\textwidth]{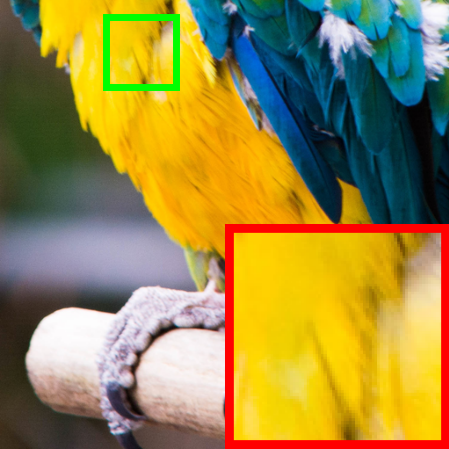}\\
\includegraphics[width=0.08\textwidth]{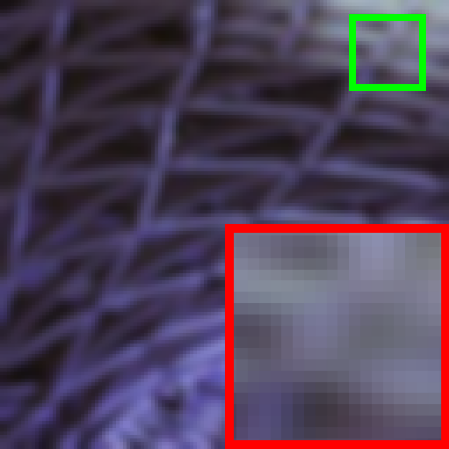}
&\includegraphics[width=0.08\textwidth]{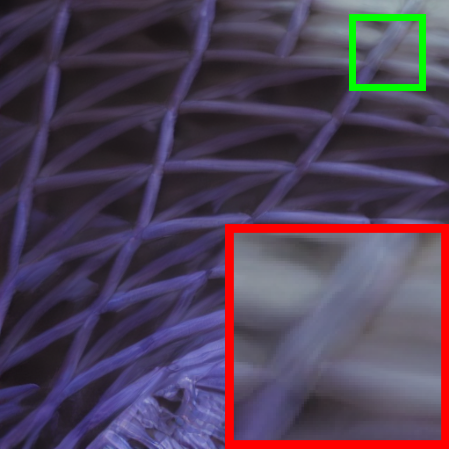}
&\includegraphics[width=0.08\textwidth]{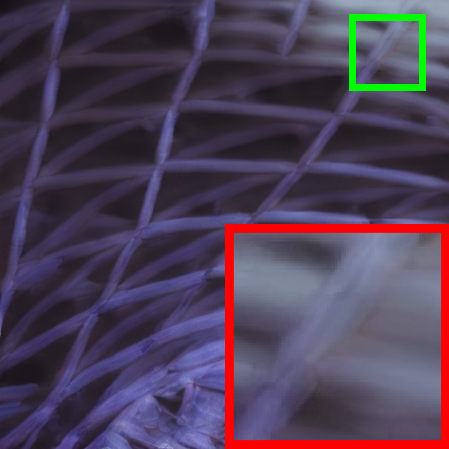}
&\includegraphics[width=0.08\textwidth]{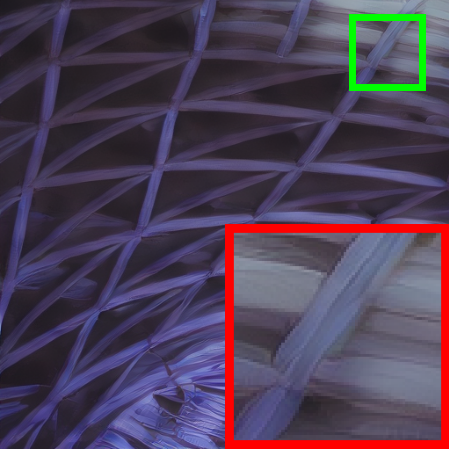}
&\includegraphics[width=0.08\textwidth]{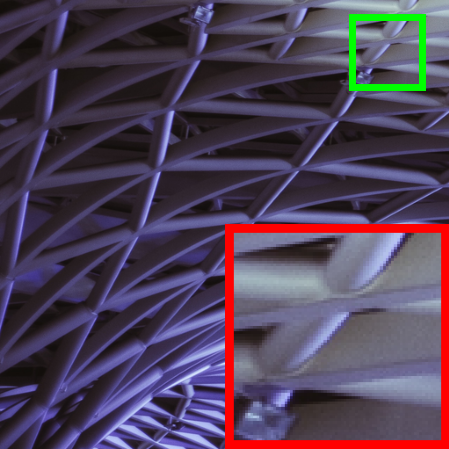}\\
\end{tabular}}
\vspace{-10pt}
\caption{\textbf{Ablation study of our two structural optimizations: eliminating VAE encoder, and optimizing the connection between the denoising UNet and the VAE decoder} on ``0886\_pch \_00025" (top) and ``0892\_pch\_00015" (bottom) from DIV2K-Val.}
\label{fig:abla_effect_of_enc_and_opt}
\vspace{-5pt}
\end{figure}

\begin{figure}[!t]
\setlength{\tabcolsep}{0.5pt}
\centering
\resizebox{\linewidth}{!}{
\scriptsize
\begin{tabular}{cccccc}
Input & w/o Elim. & \textbf{Ours} & Input & w/o Elim. & \textbf{Ours} \\
\includegraphics[width=0.07\textwidth]{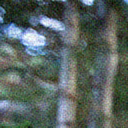}
&\includegraphics[width=0.07\textwidth]{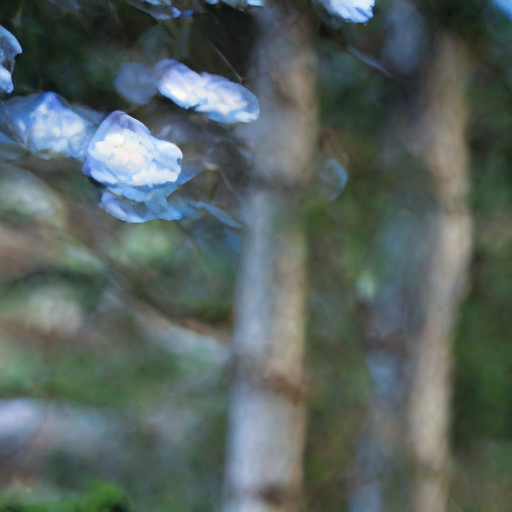}
&\includegraphics[width=0.07\textwidth]{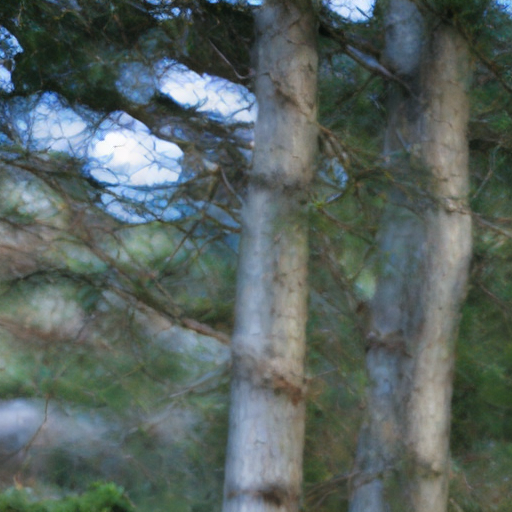}
&\includegraphics[width=0.07\textwidth]{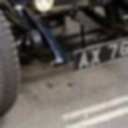}
&\includegraphics[width=0.07\textwidth]{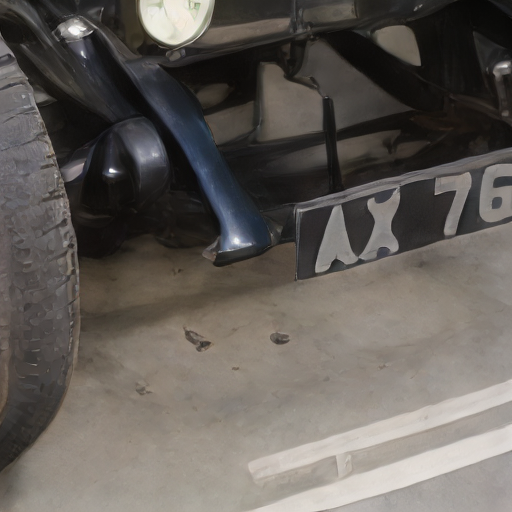}
&\includegraphics[width=0.07\textwidth]{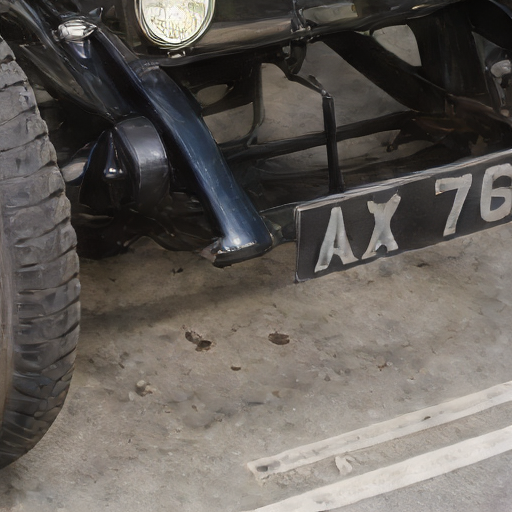}\\
\end{tabular}}
\vspace{-10pt}
\caption{\textbf{\fontsize{8.25pt}{12pt}\selectfont Ablation study of eliminating VAE encoder} on ``0815\_pch \_00001" (left) and ``0847\_pch\_00033" (right) from DIV2K-Val.}
\label{fig:abla_effect_of_enc_info}
\vspace{-5pt}
\end{figure}

\vspace{3pt}
\noindent \textbf{Efficiency Comparison.} The last 4 columns of Tab.~\ref{tab:comp_diff_quantitative} and Fig.~\ref{fig:comp_one_step} (bottom) demonstrate the superior efficiency of proposed AdcSR in step number, inference time, and computational cost. By distilling the SD-based one-step teacher \cite{wu2024one} into a structurally compressed diffusion GAN, AdcSR offers substantial speedups: 383.3$\times$, 90.7$\times$, 143.3$\times$, 93.3$\times$, and 23.7$\times$ over previous multi-step approaches StableSR, DiffBIR, SeeSR, PASD, and ResShift, respectively. Compared to the one-step model SinSR, it achieves a 4.3$\times$ acceleration. Compared to its teacher, the previously fastest method OSEDiff, it achieves a 3.7$\times$ acceleration, a 78\% reduction in computation, and a 74\% decrease in total parameters. This allows for a real-time speed of 34.79 frames per second (FPS) in diffusion-based Real-ISR. Notably, it attains a significant 9.3$\times$ speedup over S3Diff, which suffers from slower inferences due to its use of complete SD models and degradation-guided LoRA module. Fig.~\ref{fig:comp_complexity} further visualizes this efficiency comparison using a bubble plot, exhibiting the effective compression and substantial efficiency gains of AdcSR while maintaining recovery quality.

\vspace{5pt}
\noindent \fbox{\parbox{0.97\linewidth}{\quad Due to page limitations, please refer to our \Supp~for more comparison results and analyses.}}

\begin{table}[!t]
\vspace{-5pt}
\centering
\setlength{\tabcolsep}{2pt}
\caption{\textbf{Ablation study of removing the prompt extractor, text encoder, time embeddings, and related modules} on RealSR.}
\vspace{-5pt}
\resizebox{\linewidth}{!}{
\begin{tabular}{l|ccc}
\shline
Method & DISTS↓ & \#Param.↓ & Time↓ \\
\hline \hline
w/ Extractor, CA, \etc, w/ Time Embeddings & \best{0.2116} & 1311 & 0.07 \\
w/o Extractor, CA, \etc, w/ Time Embeddings & 0.2130 & 471 & \best{0.03} \\
\textbf{w/o Extr., CA, \etc, w/o Time Emb. (Ours)} & 0.2129 & \best{456} & \best{0.03} \\
\shline
\end{tabular}}
\label{tab:abla_remove_extr_and_time}
\end{table}

\begin{table}[!t]
\setlength{\tabcolsep}{2pt}
\centering
\caption{\textbf{Ablation study of pruning feature channels} on RealSR.}
\vspace{-5pt}
\resizebox{\linewidth}{!}{
\begin{tabular}{l|cccc}
\shline
Pruning Ratio (UNet / VAE Decoder) & LPIPS↓ & FID↓ & \#Param.↓ & Time↓ \\
\hline \hline
0\% / 0\% (No Channel Pruning) & 0.2883 & \best{116.74} & 839 & 0.06 \\
0\% / 50\% (Less Channel Pruning) & \best{0.2839} & 118.73 & 801 & 0.04 \\
\textbf{25\% / 50\% (Ours)} & 0.2885 & 118.41 & 456 & \best{0.03} \\
50\% / 50\% (More Channel Pruning) & 0.2897 & 125.10 & \best{210} & \best{0.03} \\
\shline
\end{tabular}}
\label{tab:abla_prune_channels}
\end{table}

\begin{figure}[!t]
\setlength{\tabcolsep}{0.5pt}
\centering
\resizebox{\linewidth}{!}{
\scriptsize
\begin{tabular}{ccccc}
Input & More Prun. & \textbf{Ours} & Less Prun. & No Prun.\\
\includegraphics[width=0.08\textwidth]{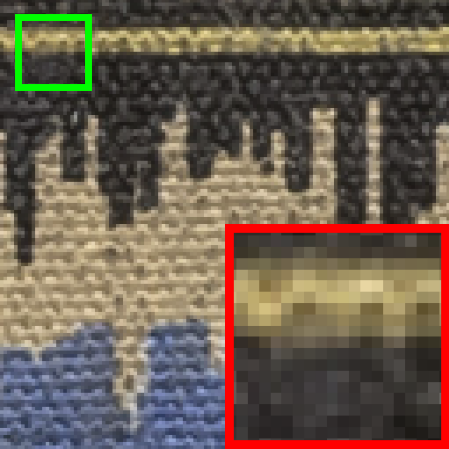}
&\includegraphics[width=0.08\textwidth]{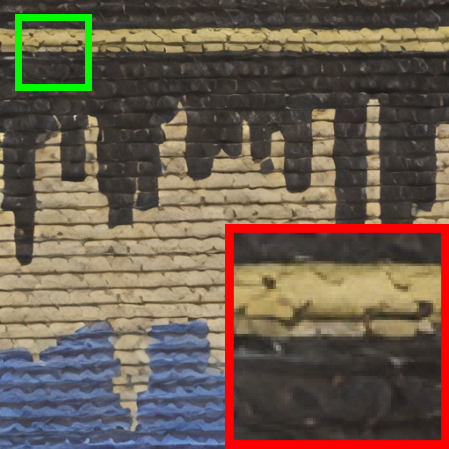}
&\includegraphics[width=0.08\textwidth]{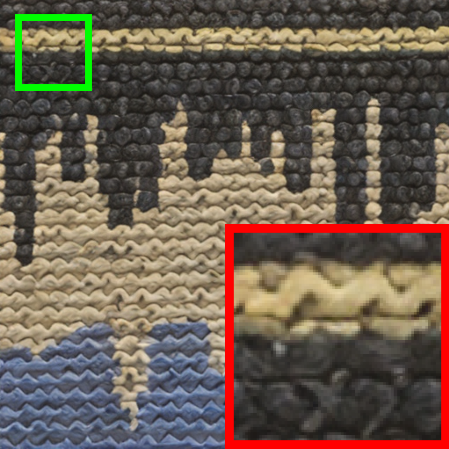}
&\includegraphics[width=0.08\textwidth]{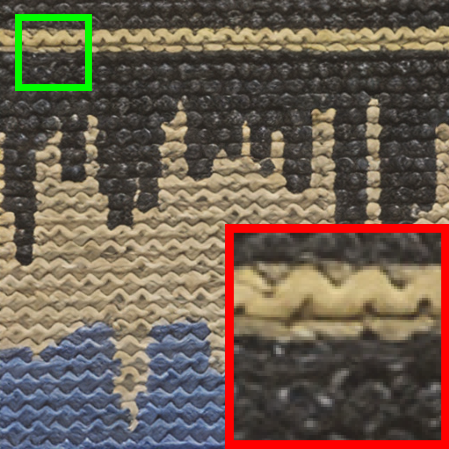}
&\includegraphics[width=0.08\textwidth]{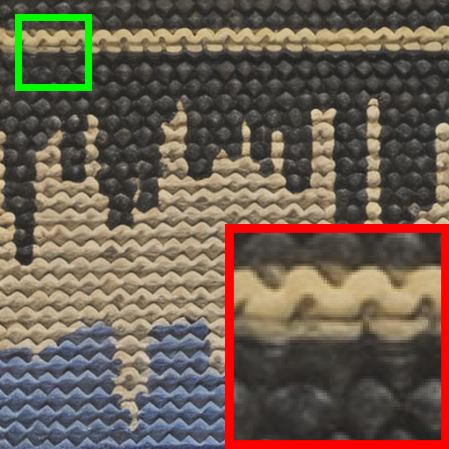}\\
\includegraphics[width=0.08\textwidth]{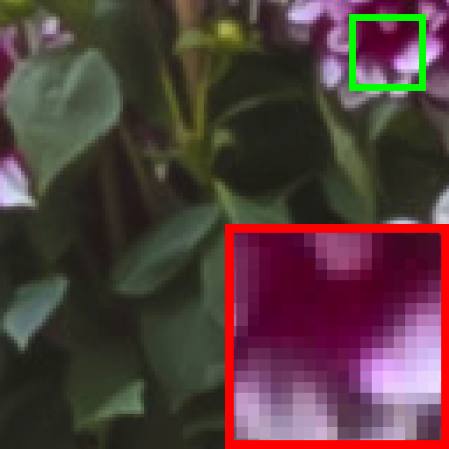}
&\includegraphics[width=0.08\textwidth]{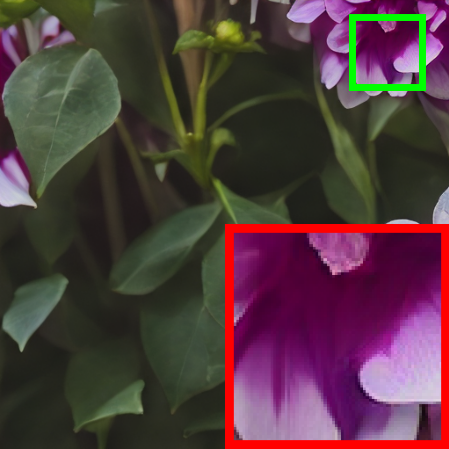}
&\includegraphics[width=0.08\textwidth]{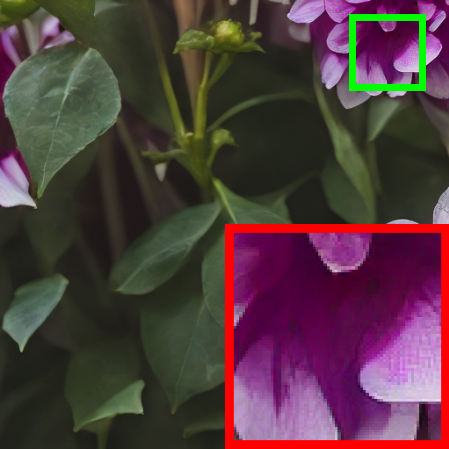}
&\includegraphics[width=0.08\textwidth]{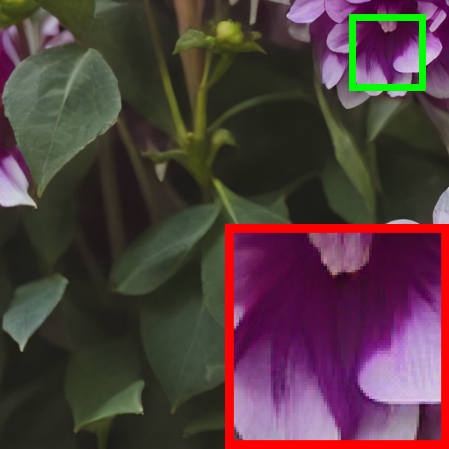}
&\includegraphics[width=0.08\textwidth]{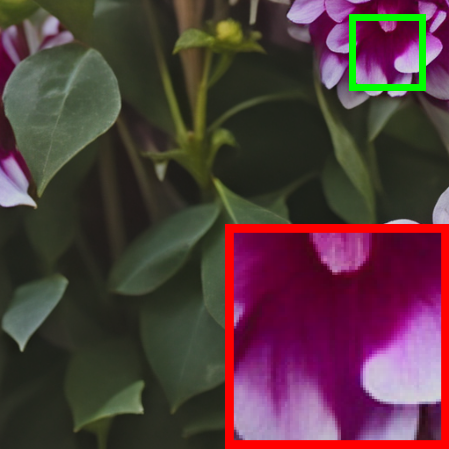}\\
UNet / Dec. & 50\% / 50\% & \textbf{25\% / 50\%} & 0\% / 50\% & 0\% / 0\%\\
\end{tabular}}
\vspace{-10pt}
\caption{\textbf{Ablation study of pruning channels} with various ratios on ``Nikon\_027" (top) and ``Nikon\_043" (bottom) from RealSR.}
\label{fig:abla_effect_of_prune}
\end{figure}

\subsection{Ablation Study}
\noindent \textbf{Effect of Eliminating the VAE Encoder, and Optimizing the UNet-VAE Decoder Connection.} Tab.~\ref{tab:abla_elim_enc} exhibits that eliminating the encoder of VAE decreases total parameter number and inference time by 9\% and 40\%, while achieving improvements of 0.13dB, 0.0031, and 0.0039 in PSNR, LPIPS, and DISTS metrics, respectively. Tab.~\ref{tab:abla_opt_conn} validates the effectiveness of optimizing the UNet-decoder connection, which brings improvements of 6.04, 1.08, 0.0120, and 0.0293 in FID, MUSIQ, MANIQA, and CLIPIQA, as well as a 0.13 FPS gain in the inference speed. Fig.~\ref{fig:abla_effect_of_enc_and_opt} visually demonstrates that omitting either of these two operations leads to noticeable blurriness in the regions of parrot’s body and the intersecting lattice beams. In particular, as exhibited in Fig.~\ref{fig:abla_effect_of_enc_info}, using the VAE encoder to compress the LR input into a latent code causes the loss of key characteristics like the clear separation between tree trunks and branches from the background, the details on the left side of the tire, the subtle shadows, and the fine textures on the car headlights. This may be attributed to the information-lossy processing of the VAE encoder. Overall, these findings indicate that directly feeding the LR image into denoising UNet, and connecting the UNet’s features before its final layer to the VAE decoder, without passing through the VAE encoder or compressing into a latent code, can effectively enhance both the fidelity and perceptual quality of super-resolved images.

\vspace{3pt}
\noindent \textbf{Effect of Removing the Text and Time Modules.} Tab.~\ref{tab:abla_remove_extr_and_time} manifests the efficiency gains brought by removing these modules. Concretely, removing extractor, text encoder, and CA layers reduces parameters by 64\% and time by 57\%, with a 0.0014 increase in DISTS. Furthermore, the removal of time embeddings results in a 0.0001 boost in DISTS and an extra 3\% reduction in parameters. Considering the significant decrease in complexity with minor recovery quality drops, these removals are incorporated into our approach.

\vspace{3pt}  
\noindent \textbf{Effect of Pruning Feature Channels.} Tab.~\ref{tab:abla_prune_channels} presents the results of channel pruning. Our method (pruning 25\% channels in the UNet and 50\% in the VAE decoder) achieves notable reductions of 46\% in parameter number and 50\% in inference time compared to the baseline (no channel pruning), with minor drops of 0.0002 in LPIPS and 1.67 in FID. However, more aggressive pruning (50\% in both UNet and VAE decoder) leads to a further 54\% reduction in parameters but results in a higher increase of 6.69 in FID and no gains in speed. Fig.~\ref{fig:abla_effect_of_prune} shows that more pruning significantly impairs the ability of model to recover textures. Therefore, we choose pruning ratios 25\% and 50\% as default settings.

\begin{table}[!t]
\setlength{\tabcolsep}{2pt}
\centering
\caption{\textbf{Ablation study of knowledge distillation} on RealSR.}
\vspace{-5pt}
\resizebox{\linewidth}{!}{
\begin{tabular}{l|ccc}
\shline
Method & PSNR↑ & NIQE↓ & CLIPIQA↑ \\
\hline \hline
{\fontsize{9.25pt}{12pt}\selectfont w/o Dist. (Replacing $\Loss_{\text{distill}}$ with $\lVert \mathbf{f}_{\text{student}} - \mathbf{f}_{\text{GT}} \rVert_{1}$)} & \best{26.75} & 8.59 & 0.5329 \\
w/ Distillation in Image Space & 25.80 & 6.74 & 0.4742 \\
w/ Dist. in Feature Space of Level 4 & 25.54 & 6.51 & 0.6168 \\
w/ Dist. in Feature Space of Level 3 & 25.49 & 5.91 & 0.6591 \\
w/ Dist. in Feature Space of Level 2 & 25.43 & 5.83 & 0.6635 \\
\textbf{w/ Dist. in Feature Space of Level 1 (Ours)} & 25.47 & \best{5.35} & \best{0.6731} \\
\shline
\end{tabular}}
\label{tab:abla_distil}
\end{table}

\begin{figure}[!t]
\setlength{\tabcolsep}{0.5pt}
\centering
\resizebox{\linewidth}{!}{
\scriptsize
\begin{tabular}{ccccc}
Input & w/o Dist. & Dist. in IS & w/o Adv. & \textbf{Ours}\\
\includegraphics[width=0.08\textwidth]{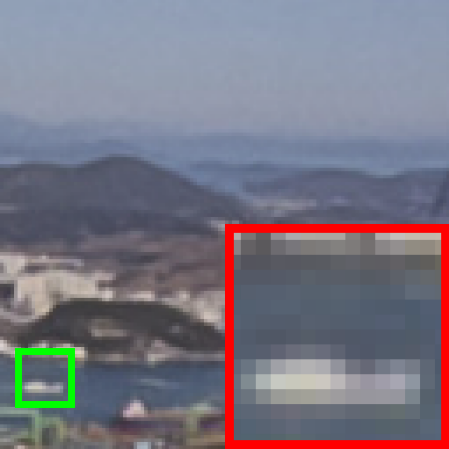}
&\includegraphics[width=0.08\textwidth]{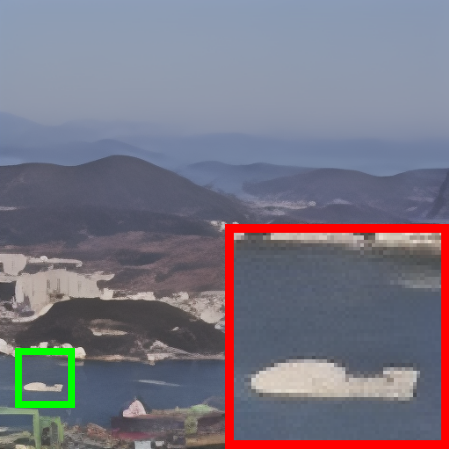}
&\includegraphics[width=0.08\textwidth]{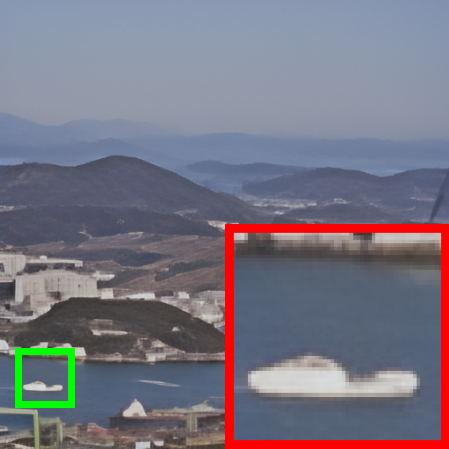}
&\includegraphics[width=0.08\textwidth]{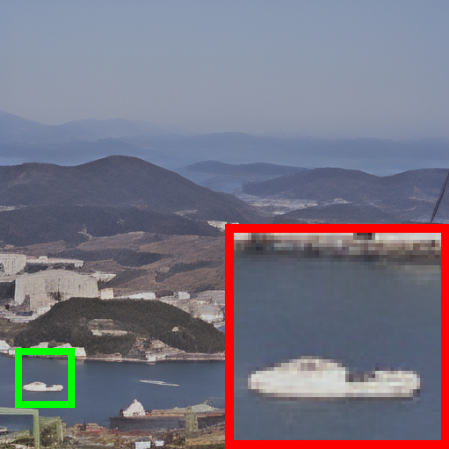}
&\includegraphics[width=0.08\textwidth]{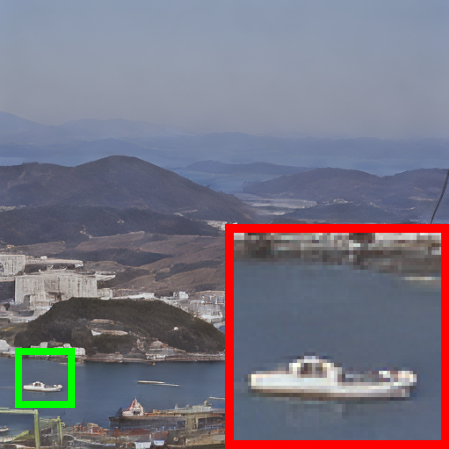}\\
\includegraphics[width=0.08\textwidth]{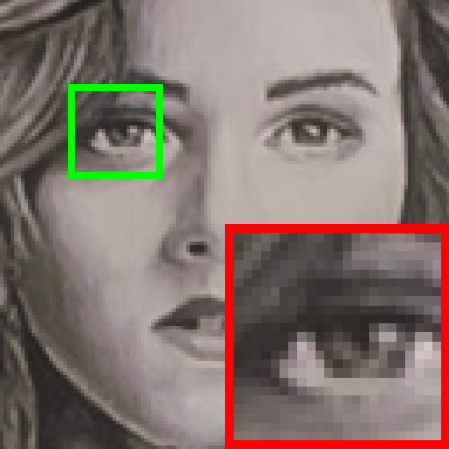}
&\includegraphics[width=0.08\textwidth]{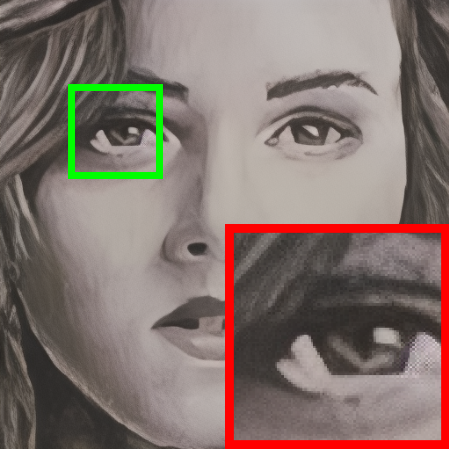}
&\includegraphics[width=0.08\textwidth]{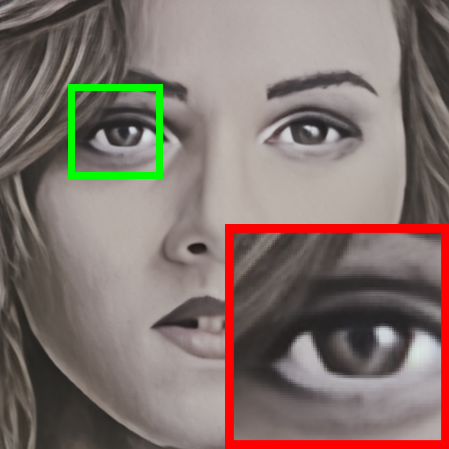}
&\includegraphics[width=0.08\textwidth]{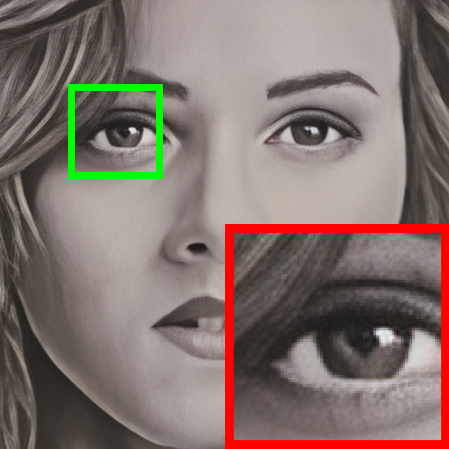}
&\includegraphics[width=0.08\textwidth]{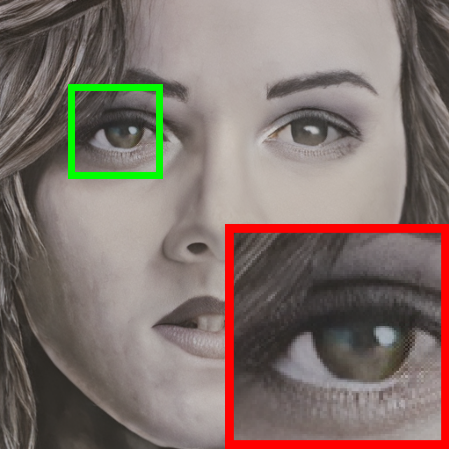}\\
\end{tabular}}
\vspace{-10pt}
\caption{\textbf{Ablation study of knowledge distillation in a feature space \textit{vs.} image space (IS) and the effect of adversarial loss} on ``Canon\_006" (top) and ``Nikon\_046" (bottom) from RealSR.}
\label{fig:abla_effect_of_distil_and_adv}
\end{figure}

\vspace{3pt}
\noindent \textbf{Effect of Knowledge Distillation in Feature Space.} Tab.~\ref{tab:abla_distil} studies the effect of distilling at various decoder levels, from level 1 (smallest spatial size) to 4 (largest size), and in image space. Firstly, we observe that replacing $\mathbf{f}_{\text{teacher}}$ with $\mathbf{f}_{\text{GT}}$ in loss $\Loss_{\text{distill}} = \lVert \mathbf{f}_{\text{student}} - \mathbf{f}_{\text{teacher}} \rVert_{1}$ significantly degrades the perceptual quality, leading to a deterioration of 3.24 in NIQE and 0.1402 in CLIPIQA. This confirms the effectiveness of knowledge distillation. Secondly, conducting distillation in the feature space with smallest spatial size achieves the best perceptual quality, yielding improvements of 1.39 in NIQE and 0.1989 in CLIPIQA compared to distillation in image space. Fig.~\ref{fig:abla_effect_of_distil_and_adv} visually demonstrates that omitting the distillation or performing it in image domain introduces distortions and blurriness in the super-resolved results, validating the effectiveness of our distillation scheme in ADC.

\vspace{3pt}
\noindent \textbf{Effect of Adversarial Loss.} Tab.~\ref{tab:abla_adv} shows the impact of various settings for $\Loss_\text{adv}$. Omitting $\Loss_\text{adv}$ significantly degrades perceptual quality by 0.0115, 0.0192, 4.12, and 0.0277 in LPIPS, DISTS, MUSIQ, and CLIPIQA, respectively. Using $\Loss_\text{adv}$ with real features $\f_\text{teacher}$ as in \cite{kang2024distilling} without leveraging $\f_\text{GT}$ results in non-negligible performance drops of 0.0162, 0.0176, 0.23, and 0.0011 in these four metrics. Fig.~\ref{fig:abla_effect_of_distil_and_adv} further illustrates that, compared to omitting $\Loss_\text{adv}$, our scheme enhances details in the boats and the woman's face, making textures in the cabin, eyelashes, and iris more natural. These results validate that our adversarial learning scheme effectively utilizes GT to improve the realism of super-resolved images, enabling the model to learn beyond its teacher.

\begin{table}[!t]
\setlength{\tabcolsep}{2pt}
\centering
\caption{\textbf{Ablation study of using adversarial loss} on DRealSR.}
\vspace{-5pt}
\resizebox{\linewidth}{!}{
\begin{tabular}{l|cccc}
\shline
Method & LPIPS↓ & DISTS↓ & MUSIQ↑ & MANIQA↑ \\
\hline \hline
w/o Adversarial Loss & 0.3261 & 0.2392 & 62.14 & 0.5650 \\
w/ Adv. (Real Feature: $\f_\text{teacher}$) & 0.3208 & 0.2376 & 66.03 & 0.5916 \\
\textbf{w/ Adv. (Real Feat.: $\f_\text{GT}$) (Ours)} & \best{0.3046} & \best{0.2200} & \best{66.26} & \best{0.5927} \\
\shline
\end{tabular}}
\label{tab:abla_adv}
\end{table}

\section{Conclusion}
In this paper, we proposed a novel method, \textbf{AdcSR}, based on our \textbf{A}dversarial \textbf{D}iffusion \textbf{C}ompression (\textbf{ADC}) framework, for real-world image super-resolution (Real-ISR). To be specific, we structurally compressed a typical state-of-the-art SD-based one-step diffusion network, OSEDiff, into a smaller diffusion GAN. We identified and removed unnecessary modules (VAE encoder, prompt extractor, \etc) from OSEDiff, and pruned its remaining compressible modules (denoising UNet and VAE decoder). Since direct removal and pruning can degrade the model's generative capability, we developed a two-stage training scheme that first pretrains a pruned SD VAE decoder and then performs adversarial distillation to compensate for performance loss. Experiments on both synthetic and real-world datasets demonstrated that our AdcSR model delivered competitive image quality and superior computational efficiency compared to existing diffusion-based Real-ISR approaches.

While ADC and AdcSR have demonstrated effectiveness in compressing SD-based one-step Real-ISR network and achieving real-time inference, they face challenges in accurately recovering fine textures and heavily degraded details, as shown in Fig.~\ref{fig:abla_effect_of_enc_and_opt}. Moreover, although this work focuses on streamlining the state-of-the-art Real-ISR model OSEDiff, our ADC framework could be extended to other SD-based methods. We plan to explore such extensions and integrate additional generative priors for Real-ISR in future work.

\appendix

\renewcommand{\thesection}{\Alph{section}}
\counterwithin{figure}{section}
\counterwithin{table}{section}

\section*{Supplementary Material}
Our main paper outlines the core idea and techniques of proposed method. It also demonstrates the effectiveness of our four main methodological contributions and adopted settings through experimental validation. In this \Supp, we provide additional details, including the training and inference pseudocode of proposed ADC framework in Sec.~\ref{sec:pseudocode_of_algo}, more ablation studies in Sec.~\ref{sec:more_ablation_studies}, more comparison results and a user study analysis in Sec.~\ref{sec:more_comparison_results}, as well as an efficiency evaluation of AdcSR and its SD-based one-step teacher OSEDiff \cite{wu2024one} on a real mobile platform, which are not included in the main paper due to space constraints.

\section{Pseudocode of Training and Inference}
\label{sec:pseudocode_of_algo}
In this section, we present the training and inference procedures of our ADC framework, as summarized in Algo.~\ref{alg}. The training process consists of two stages: \textbf{(1)} pretraining the channel-pruned SD VAE decoder to restore its decoding ability, and \textbf{(2)} knowledge distillation with adversarial loss to compensate for performance degradation due to our compression. The inference of AdcSR is faster than complete SD \cite{rombach2022high,stabilityai} models due to its compressed structure.

\section{More Ablation Studies}
\label{sec:more_ablation_studies}
\noindent \textbf{Effect of Channel Pruning.} Tab.~\ref{tab:abla_compression_strategy} compares employed channel pruning to other two alternative structural compression strategies: using a block-removed UNet \cite{kim2023bk,bksdmv2small} and decoding by the pretrained tiny VAE \cite{dao2024swiftbrush,tinyvae}. We observe that, with similar parameter numbers and inference speed, applying block removal results in a noticeable performance loss of 0.0083 and 0.0084 in LPIPS and DISTS, respectively. While the use of tiny VAE decoder can lead to reductions of 12M parameters and 0.01s in inference time, it substantially degrades performance by 0.0297 and 0.0161 in LPIPS and DISTS. This may be attributed to the reduced depth and the absence of global receptive field in tiny VAE, which relies solely on convolutions for decoding. These results validate the effectiveness of our adopted feature channel pruning.

\vspace{3pt}
\noindent \textbf{Effect of Various LoRA Ranks, and Fully Finetuning the First Layer for the Discriminator.} Tab.~\ref{tab:abla_lora_and_fully_finetune_first_layer} compares various finetuning settings for discriminator. Fully finetuning it can lead to unstable training without convergence. Compared to the rank of 2, a rank of 4 achieves notable quality gains of 0.0009, 2.09, 6.58, and 0.0139 in evaluation metrics DISTS, FID, MUSIQ, and CLIPIQA, respectively. In contrast, higher ranks of 8 and 16 bring no evident improvements. Furthermore, based on the rank of 4, fully finetuning the first convolution layer further enhances performance by 0.0042, 1.26, 1.05, and 0.0536 in these four metrics. These results validate the effectiveness of our default ADC setting.

\begin{table}[!t]
\setlength{\tabcolsep}{3pt}
\centering
\caption{\textbf{\fontsize{8.5pt}{12pt}\selectfont Ablation study of structural compression} on DRealSR.}
\vspace{-5pt}
\resizebox{\linewidth}{!}{
\begin{tabular}{l|cccc}
\shline
Method (UNet / VAE Decoder) & LPIPS↓ & DISTS↓ & \#Param.↓ & Time↓ \\
\hline \hline
Block-Removed / Channel-Pruned & 0.3129 & 0.2284 & 457 & 0.03 \\
Channel-Pruned / Tiny VAE Dec. & 0.3343 & 0.2361 & \best{444} & \best{0.02} \\
\textbf{Ch.-Pruned / Ch.-Pruned (Ours)} & \best{0.3046} & \best{0.2200} & 456 & 0.03 \\
\shline
\end{tabular}}
\label{tab:abla_compression_strategy}
\end{table}

\begin{table}[!t]
\setlength{\tabcolsep}{2pt}
\centering
\caption{\textbf{Ablation study of LoRA rank $r$ and fully finetuning (FT.) the first convolution layer for discriminator} on RealSR.}
\vspace{-5pt}
\resizebox{\linewidth}{!}{
\begin{tabular}{l|cccc}
\shline
Method & DISTS↓ & FID↓ & MUSIQ↑ & CLIPIQA↑ \\
\hline \hline
Fully Finetuning the Discriminator & \multicolumn{4}{c}{- (No Convergence)} \\
$r=2$ (w/o Fully FT. the 1st Layer) & 0.2182 & 121.76 & 62.27 & 0.6056 \\
$r=4$ (w/o Fully FT. the 1st Layer) & 0.2171 & 119.67 & 68.85 & 0.6195 \\
$r=8$ (w/o Fully FT. the 1st Layer) & 0.2182 & 120.94 & 68.76 & 0.6114 \\
$r=16$ (w/o Fully FT. the 1st Layer) & 0.2191 & 120.33 & 68.72 & 0.6173 \\
\textbf{$r=4$ (w/ Fully FT. 1st Lyr.) (Ours)} & \best{0.2129} & \best{118.41} & \best{69.90} & \best{0.6731} \\
\shline
\end{tabular}}
\label{tab:abla_lora_and_fully_finetune_first_layer}
\end{table}

\begin{algorithm}[!t]
\caption{Training and Inference of ADC}
\label{alg}
~\\
\KwIn{Pretrained one-step teacher; Pretrained SD models: VAE encoder $\Enc_\text{SD}$, VAE decoder $\Dec_\text{SD}$, UNet $\Ep_{\text{SD}}$; Weighting factor $\lambda_\text{adv}$.}
\BlankLine
\textbf{Stage 1: Pretraining Pruned VAE Decoder}\\
Prune the SD VAE decoder $\mathcal{D}_{\text{SD}}$ to obtain $\Dec_\text{pruned}$\;
Initialize $\Dec_\text{pruned}$ and a discriminator as in \cite{rombach2022high,ldm}\;
\For{number of training iterations}{
Sample a batch of images $\x$\;
Obtain latent codes $\z = \Enc_\text{SD}(\x)$\;
Reconstruct images $\xhat = \Dec_{\text{pruned}}(\z)$\;
Compute reconstruction loss \cite{rombach2022high} of $\x$ and $\xhat$\;
Update $\Dec_\text{pruned}$ using Adam optimizer\;
Compute discriminator loss \cite{rombach2022high} of $\x$ and $\xhat$\;
Update discriminator using Adam optimizer\;
}
\BlankLine
~\\
\textbf{Stage 2: Adversarial Distillation}\\
Prune the SD UNet $\Ep_\text{SD}$ to obtain $\Ep_\text{pruned}$\;
Initialize the student model using $\Ep_\text{pruned}$ and $\Dec_\text{pruned}$\;
Initialize a feature-space discriminator using $\Ep_\text{SD}$\;
\For{number of training iterations}{
Sample a batch of LR-HR pairs $(\x_{\text{LR}}, \x_\text{HR})$\;
Compute features $\f_\text{student}$ from student model\;
Compute features $\f_\text{teacher}$ from teacher model\;
Compute distillation loss:
$$
\Loss_\text{distill} = \lVert \f_\text{student} - \f_\text{teacher} \rVert_1
$$ \\
Compute adversarial loss:
$$
\Loss_\text{adv} = \text{Softplus} \left( -\text{Discriminator}(\f_\text{student}) \right)
$$ \\
Compute total loss: $\Loss = \Loss_\text{distill} + \lambda_\text{adv} \Loss_\text{adv}$\;
Update student model using Adam optimizer\;
Compute features $\f_\text{GT}$ using $\x_\text{HR}$\;
Compute discriminator loss:
\begin{align*}
\Loss_\text{disc} =~&\text{Softplus} \left(\text{Discriminator}(\f_\text{student}) \right)\\
+~&\text{Softplus} \left( -\text{Discriminator}(\f_\text{GT}) \right)
\end{align*} \\
Update discriminator using Adam optimizer\;
}
\BlankLine
~\\
\textbf{Inference}\;
Given LR image input $\x_\text{LR}$\;
\Return Super-resolved image $\xhat_\text{HR} = \text{Student}(\x_\text{LR})$\;
\end{algorithm}

\section{More Comparison Results on Benchmarks}
\label{sec:more_comparison_results}
\subsection{More Quantitative Comparisons}
In Tab.~\ref{tab:comp_quantitative}, we compare the proposed AdcSR model against twelve state-of-the-arts, including four representative GAN-based approaches: BSRGAN \cite{zhang2021designing}, Real-ESRGAN \cite{wang2021real}, LDL \cite{liang2022details}, and FeMASR \cite{chen2022real}, as well as eight diffusion-based methods \cite{wang2024exploiting,lin2023diffbir,wu2024seesr,yang2023pixel,yue2024resshift,wang2024sinsr,wu2024one,zhang2024degradation} across three synthetic and real-world test datasets, evaluated using nine metrics \cite{wang2004image,zhang2018unreasonable,ding2020image,heusel2017gans,zhang2015feature,ke2021musiq,yang2022maniqa,wang2023exploring}. We observe that, firstly, the traditional GAN-based approaches generally perform well on reference-based metrics, particularly the fidelity measures PSNR and SSIM. Secondly, diffusion-based methods outperform these GANs in most perceptual quality metrics, showing their ability to better generate natural textures. Thirdly, AdcSR achieves competitive results, surpassing its teacher OSEDiff in most cases, which validates the effectiveness of ADC's compression and adversarial distillation.

\subsection{More Qualitative Comparisons}
\begin{figure*}[!t]
\setlength{\tabcolsep}{0.5pt}
\centering
\resizebox{\linewidth}{!}{
\tiny
\begin{tabular}{ccccccc}
Input & BSRGAN & Real-ESRGAN & LDL & FeMASR & StableSR & DiffBIR \\
\includegraphics[width=0.08\textwidth]{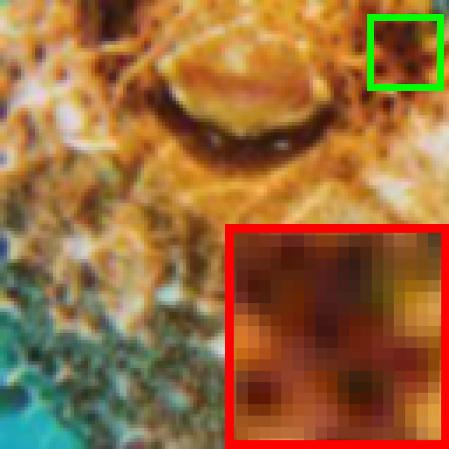}
&\includegraphics[width=0.08\textwidth]{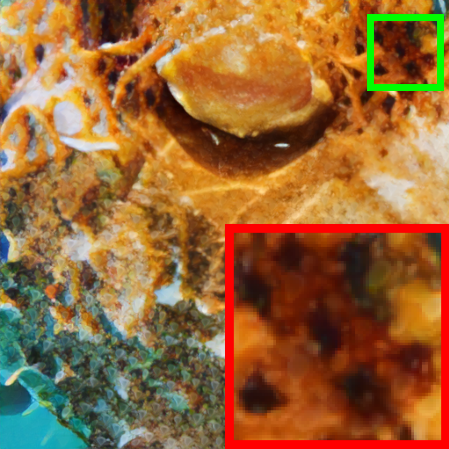}
&\includegraphics[width=0.08\textwidth]{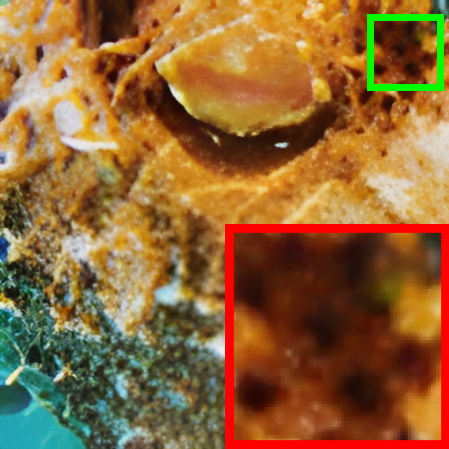}
&\includegraphics[width=0.08\textwidth]{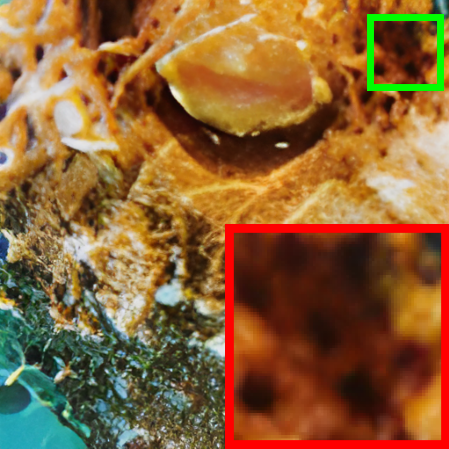}
&\includegraphics[width=0.08\textwidth]{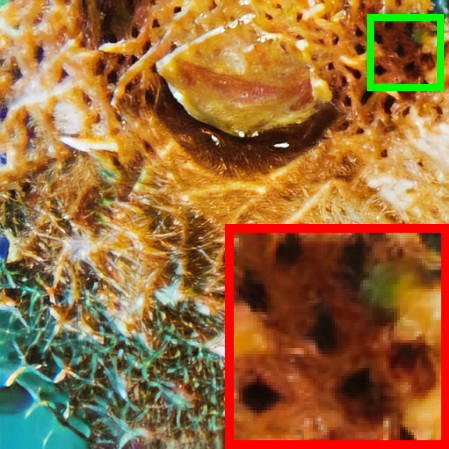}
&\includegraphics[width=0.08\textwidth]{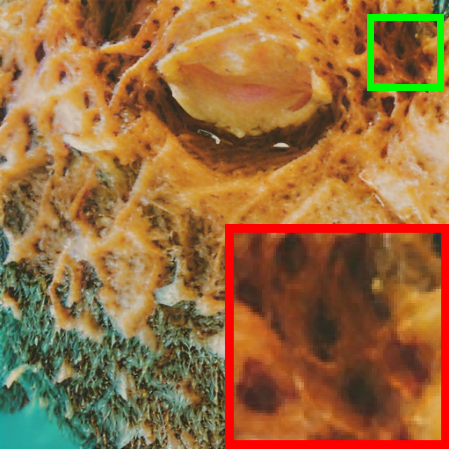}
&\includegraphics[width=0.08\textwidth]{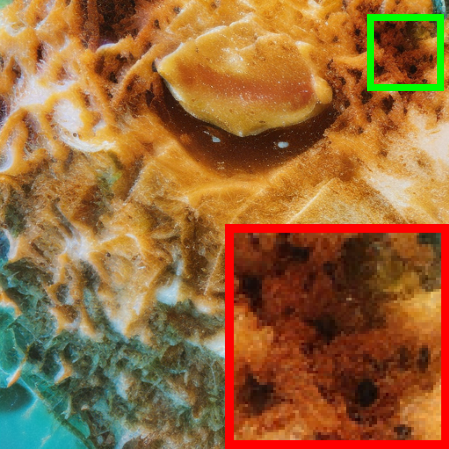}\\
SeeSR & PASD & ResShift & SinSR & OSEDiff & S3Diff & \textbf{AdcSR (Ours)}\\
\includegraphics[width=0.08\textwidth]{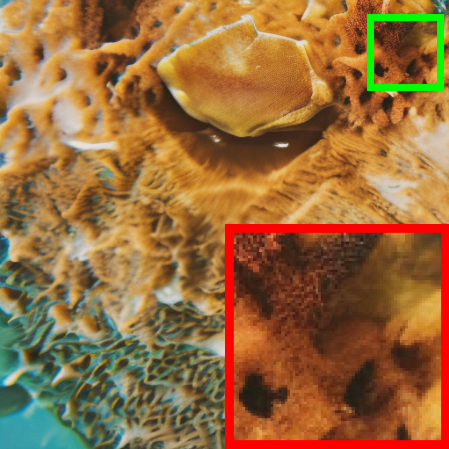}
&\includegraphics[width=0.08\textwidth]{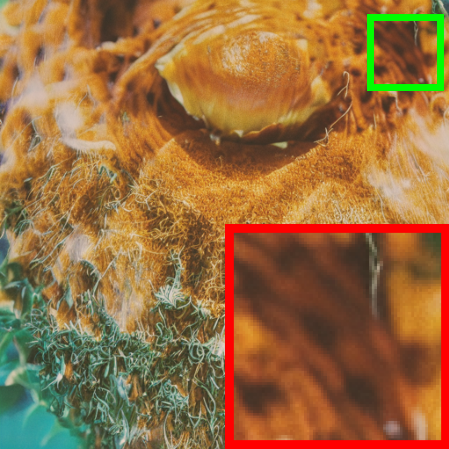}
&\includegraphics[width=0.08\textwidth]{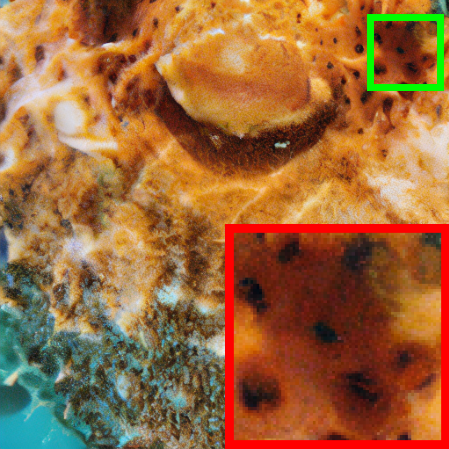}
&\includegraphics[width=0.08\textwidth]{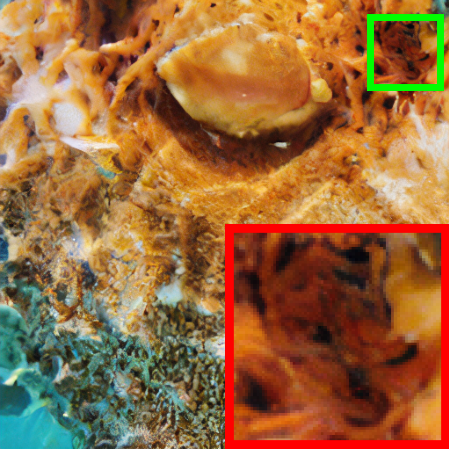}
&\includegraphics[width=0.08\textwidth]{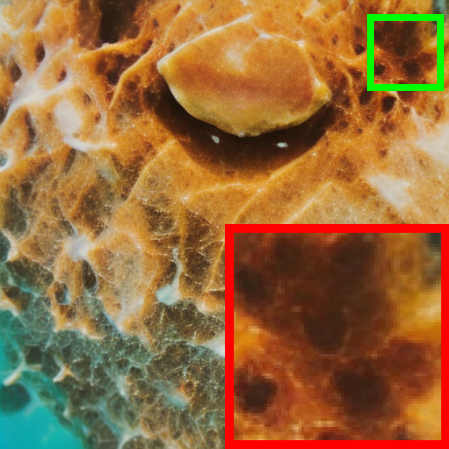}
&\includegraphics[width=0.08\textwidth]{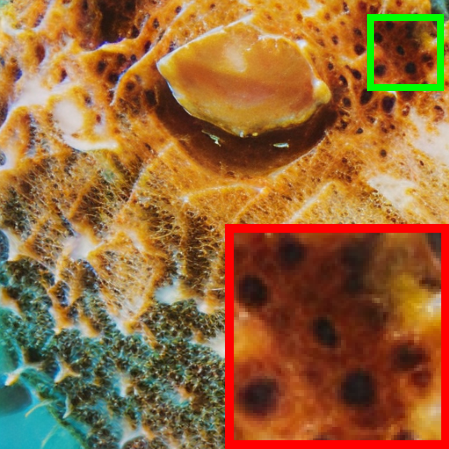}
&\includegraphics[width=0.08\textwidth]{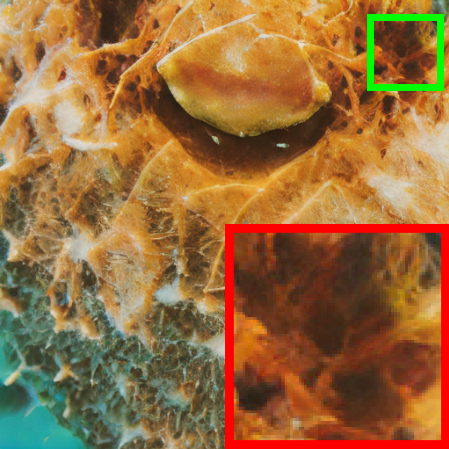}\\
\end{tabular}}
\vspace{-10pt}
\caption{\textbf{Qualitative comparison of different approaches} on an image named ``0835\_pch\_00017" from the DIV2K-Val \cite{wang2024exploiting} dataset.}
\label{fig:comp_qualitative_DIV2K_Val}
\end{figure*}

\begin{figure*}[!t]
\setlength{\tabcolsep}{0.5pt}
\centering
\resizebox{\linewidth}{!}{
\tiny
\begin{tabular}{ccccccc}
Input & BSRGAN & Real-ESRGAN & LDL & FeMASR & StableSR & DiffBIR \\
\includegraphics[width=0.08\textwidth]{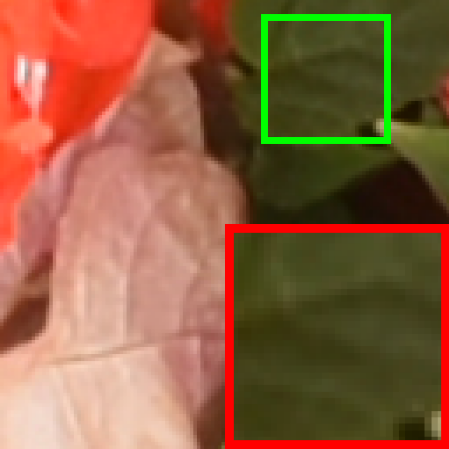}
&\includegraphics[width=0.08\textwidth]{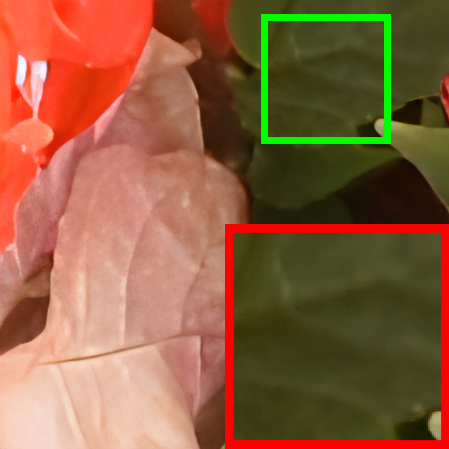}
&\includegraphics[width=0.08\textwidth]{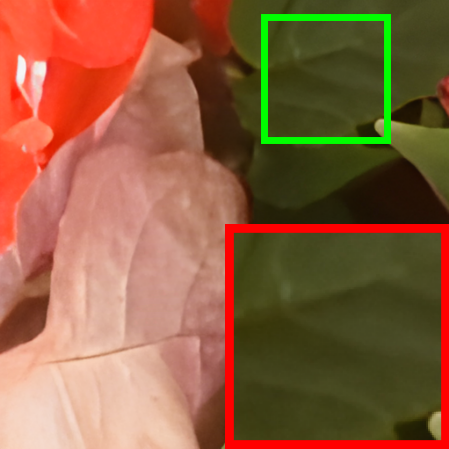}
&\includegraphics[width=0.08\textwidth]{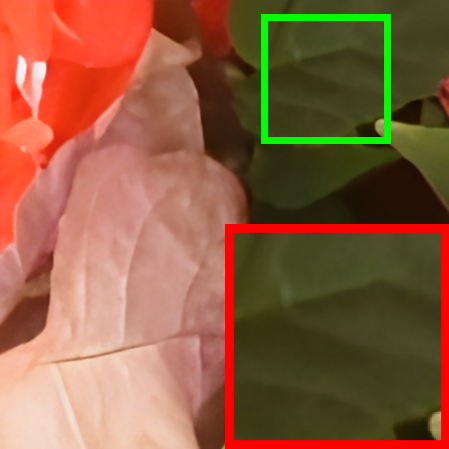}
&\includegraphics[width=0.08\textwidth]{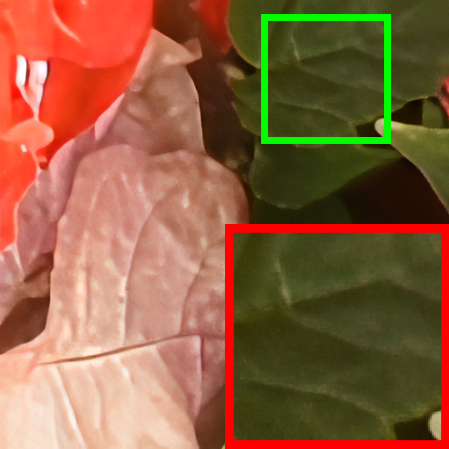}
&\includegraphics[width=0.08\textwidth]{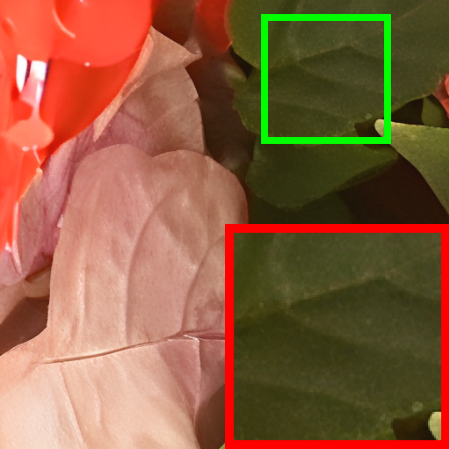}
&\includegraphics[width=0.08\textwidth]{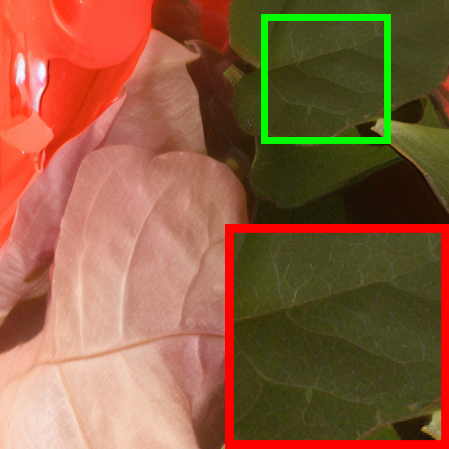}\\
SeeSR & PASD & ResShift & SinSR & OSEDiff & S3Diff & \textbf{AdcSR (Ours)}\\
\includegraphics[width=0.08\textwidth]{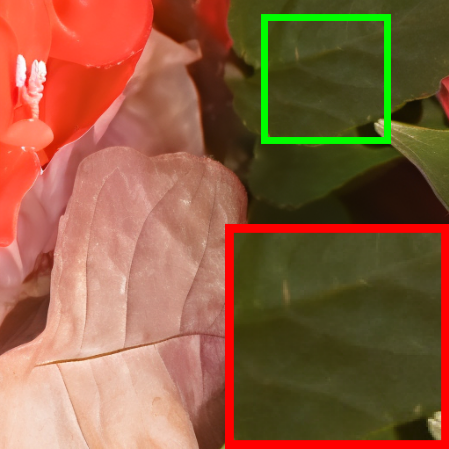}
&\includegraphics[width=0.08\textwidth]{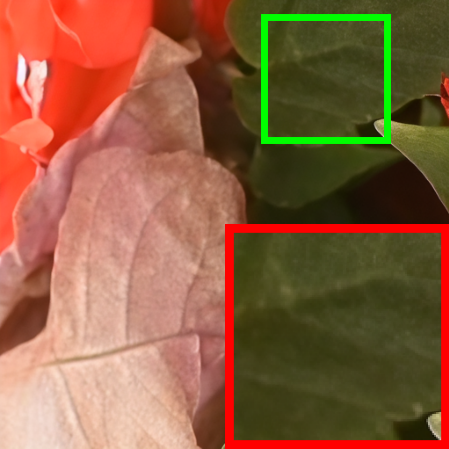}
&\includegraphics[width=0.08\textwidth]{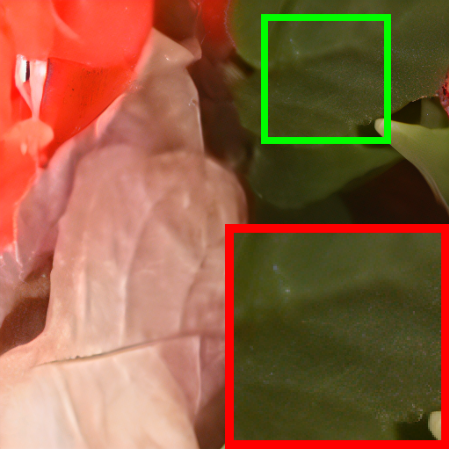}
&\includegraphics[width=0.08\textwidth]{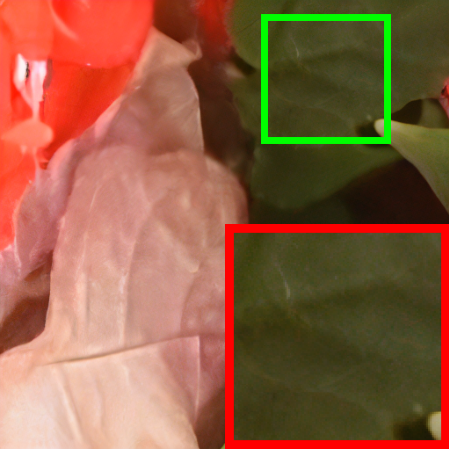}
&\includegraphics[width=0.08\textwidth]{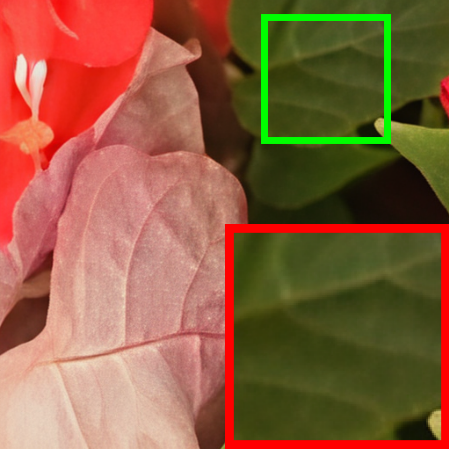}
&\includegraphics[width=0.08\textwidth]{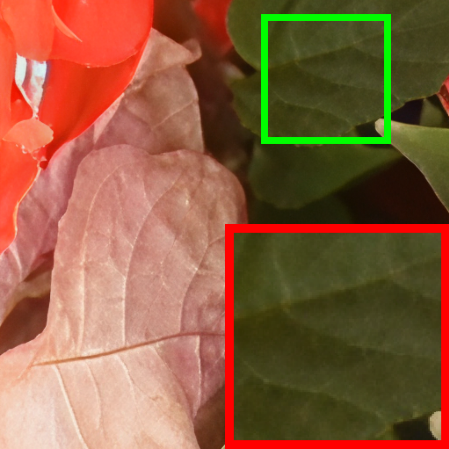}
&\includegraphics[width=0.08\textwidth]{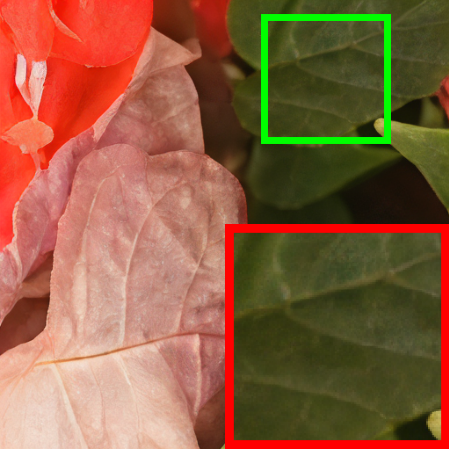}\\
\end{tabular}}
\vspace{-10pt}
\caption{\textbf{Qualitative comparison of different approaches} on a real-world image named ``DSC\_1599" from the DRealSR \cite{wei2020component} dataset.}
\label{fig:comp_qualitative_DRealSR}
\end{figure*}

\begin{figure*}[!t]
\setlength{\tabcolsep}{0.5pt}
\centering
\resizebox{\linewidth}{!}{
\tiny
\begin{tabular}{ccccccc}
Input & BSRGAN & Real-ESRGAN & LDL & FeMASR & StableSR & DiffBIR \\
\includegraphics[width=0.08\textwidth]{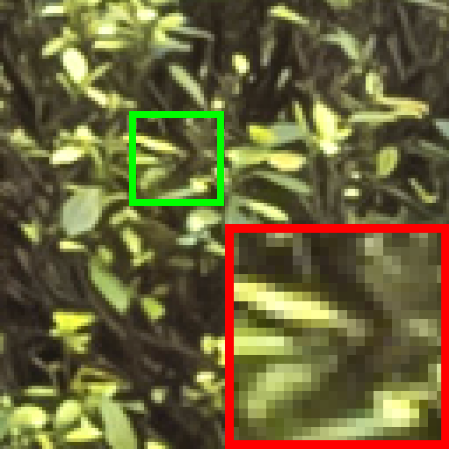}
&\includegraphics[width=0.08\textwidth]{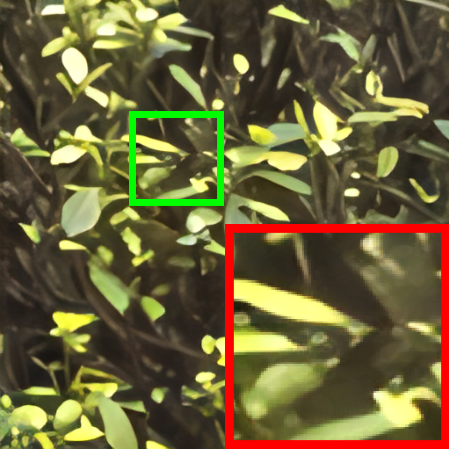}
&\includegraphics[width=0.08\textwidth]{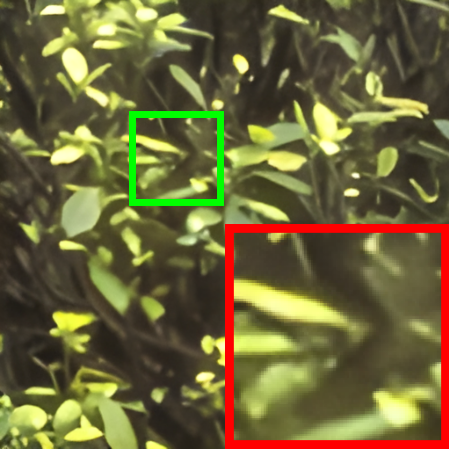}
&\includegraphics[width=0.08\textwidth]{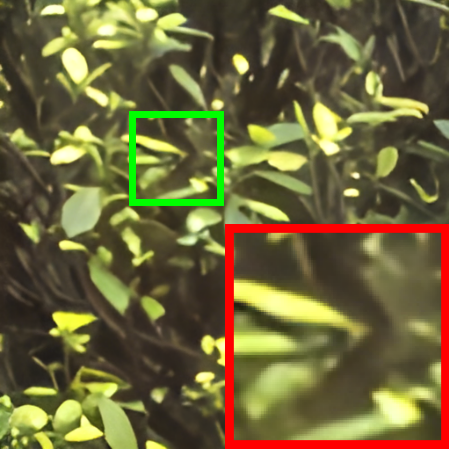}
&\includegraphics[width=0.08\textwidth]{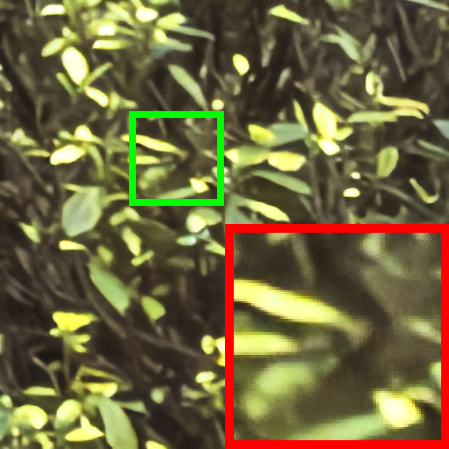}
&\includegraphics[width=0.08\textwidth]{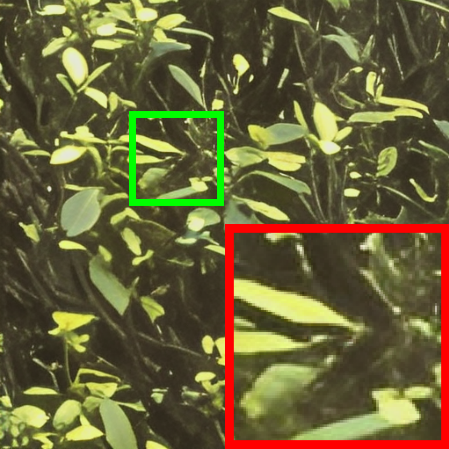}
&\includegraphics[width=0.08\textwidth]{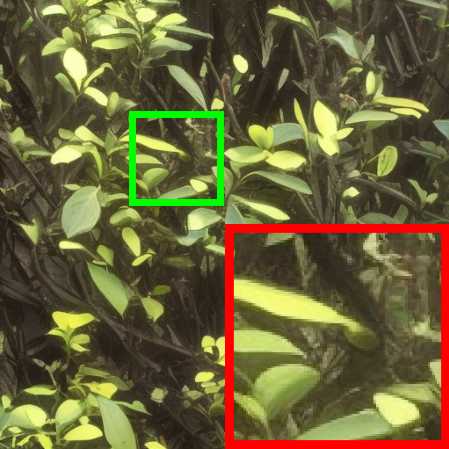}\\
SeeSR & PASD & ResShift & SinSR & OSEDiff & S3Diff & \textbf{AdcSR (Ours)}\\
\includegraphics[width=0.08\textwidth]{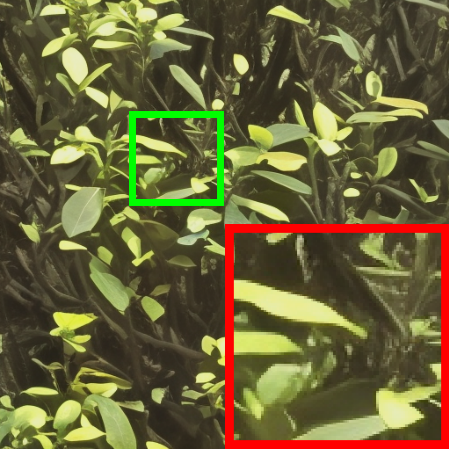}
&\includegraphics[width=0.08\textwidth]{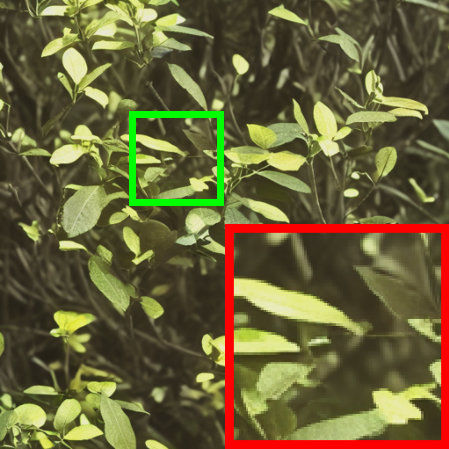}
&\includegraphics[width=0.08\textwidth]{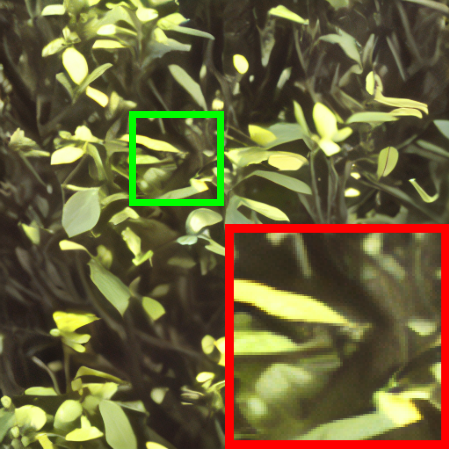}
&\includegraphics[width=0.08\textwidth]{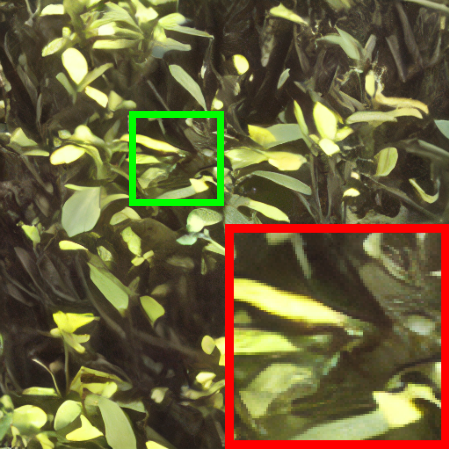}
&\includegraphics[width=0.08\textwidth]{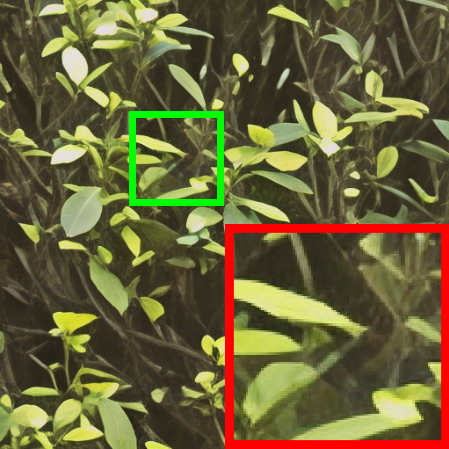}
&\includegraphics[width=0.08\textwidth]{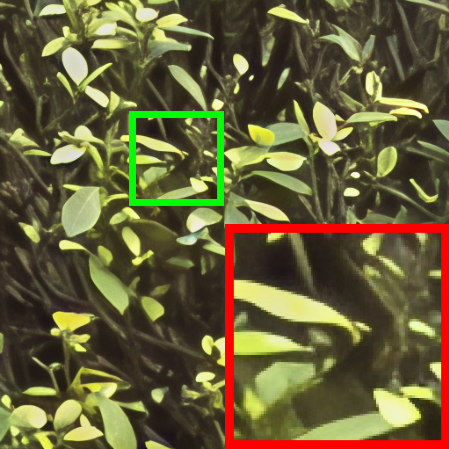}
&\includegraphics[width=0.08\textwidth]{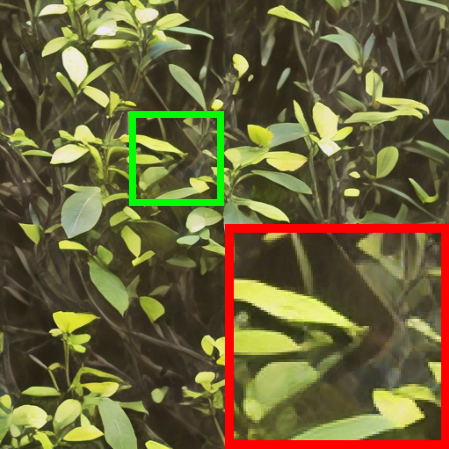}\\
\end{tabular}}
\vspace{-10pt}
\caption{\textbf{Qualitative comparison of different approaches} on a real-world image named ``Nikon\_013" from the RealSR \cite{cai2019toward} dataset.}
\label{fig:comp_qualitative_RealSR}
\end{figure*}

Figs.~\ref{fig:comp_qualitative_DIV2K_Val}, \ref{fig:comp_qualitative_DRealSR}, and \ref{fig:comp_qualitative_RealSR} present visual comparisons across super-resolution images produced by these approaches. We observe that, firstly, GAN-based approaches generally show weaker generative capabilities than diffusion-based methods, recovering fewer details overall. Secondly, traditional multi-step SD-based methods generate rich details but may introduce artifacts, such as those observed on the spiky texture of the inflated pufferfish by StableSR, DiffBIR, SeeSR, and PASD. Thirdly, ResShift and SinSR tend to produce oversmoothed results in areas of the leaves and red flower petals, where the vein structures and textures are less distinct. This may be due to their lack of exploiting the powerful SD priors. Fourthly, AdcSR demonstrates competitive performance, generating natural and balanced details in the pufferfish and leaves, comparable to OSEDiff and S3Diff, which can subtly introduce an additional slight highlight effect on the cluster of leaves. These results comprehensively confirm the effectiveness of our approach in compressing SD-based models for Real-ISR while maintaining quality.

\begin{figure}[!t]
\setlength{\tabcolsep}{1pt}
\centering
\resizebox{\linewidth}{!}{
\tiny
\begin{tabular}{cccccccc}
\includegraphics[width=0.08\textwidth]{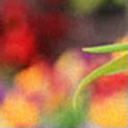}
&\includegraphics[width=0.08\textwidth]{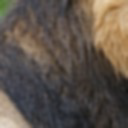}
&\includegraphics[width=0.08\textwidth]{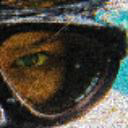}
&\includegraphics[width=0.08\textwidth]{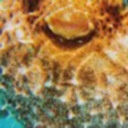}
&\includegraphics[width=0.08\textwidth]{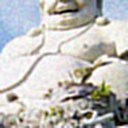}
&\includegraphics[width=0.08\textwidth]{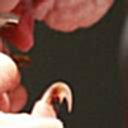}
&\includegraphics[width=0.08\textwidth]{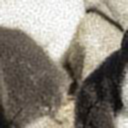}
&\includegraphics[width=0.08\textwidth]{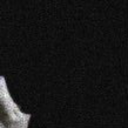}\\
\includegraphics[width=0.08\textwidth]{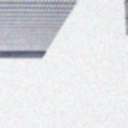}
&\includegraphics[width=0.08\textwidth]{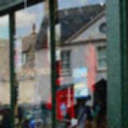}
&\includegraphics[width=0.08\textwidth]{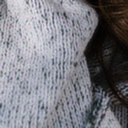}
&\includegraphics[width=0.08\textwidth]{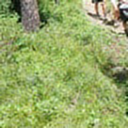}
&\includegraphics[width=0.08\textwidth]{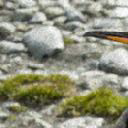}
&\includegraphics[width=0.08\textwidth]{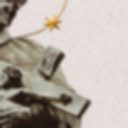}
&\includegraphics[width=0.08\textwidth]{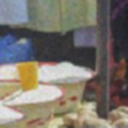}
&\includegraphics[width=0.08\textwidth]{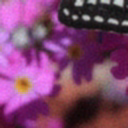}\\
\end{tabular}}
\vspace{-10pt}
\caption{\textbf{16 LR images} from DIV2K-Val adopted in user study.}
\label{fig:user_study_LR_images}
\end{figure}

\begin{table}[!t]
\centering
\caption{\textbf{\fontsize{8.5pt}{12pt}\selectfont User study results} of one-step diffusion-based methods.}
\vspace{-5pt}
\resizebox{\linewidth}{!}{
\begin{tabular}{l|cccc}
\shline
Method & SinSR & OSEDiff & S3Diff & \textbf{AdcSR (Ours)} \\
\hline \hline
Total Votes & 35 & \tbest{149} & \best{168} & \sbest{160} \\
Voting Rate (\%) & 7 & \tbest{29} & \best{33} & \sbest{31} \\
\shline
\end{tabular}}
\label{tab:user_study_voting_rate}
\end{table}

\begin{table*}[!t]
\vspace{-5pt}
\centering
\caption{\textbf{Quantitative comparison among thirteen different GAN-based and diffusion-based Real-ISR approaches on both synthetic and real-world benchmarks.} ``S'' denotes the required number of sampling steps in the diffusion-based method.}
\vspace{-5pt}
\resizebox{\linewidth}{!}{
\begin{tabular}{l|l|ccccccccc}
\shline
Test Dataset & Method & PSNR↑ & SSIM↑ & LPIPS↓ & DISTS↓ & FID↓ & NIQE↓ & MUSIQ↑ & MANIQA↑ & CLIPIQA↑ \\
\hline \hline
\multirow{13}{*}{DIV2K-Val} & BSRGAN & \sbest{24.58} & \tbest{0.6269} & 0.3351 & 0.2275 & 44.23  & 4.75 & 61.20  & 0.5071  & 0.5247   \\
& Real-ESRGAN & 24.29 & \best{0.6371} & 0.3112 & 0.2141 & 37.64 & \sbest{4.68} & 61.06 & 0.5501 & 0.5277 \\
& LDL & 23.83 & \sbest{0.6344} & 0.3256 & 0.2227 & 42.29 & 4.86 & 60.04  & 0.5350 & 0.5180 \\
& FeMASR & 23.06 & 0.5887 & 0.3126 & 0.2057 & 35.87 & 4.74 & 60.83 & 0.5074 & 0.5997 \\
\cline{2-11}
& StableSR-S200 & 23.26 & 0.5726 & 0.3113 & 0.2048 & \sbest{24.44} & 4.76 & 65.92 & 0.6192 & 0.6771 \\
& DiffBIR-S50 & 23.64 & 0.5647 & 0.3524 & 0.2128 & 30.72 & \tbest{4.70} & 65.81 & 0.6210 & 0.6704 \\
& SeeSR-S50 & 23.68 & 0.6043 & 0.3194 & \tbest{0.1968} & 25.90 & 4.81 & \sbest{68.67} & \tbest{0.6240} & \sbest{0.6936} \\
& PASD-S20 & 23.14 & 0.5505 & 0.3571 & 0.2207 & 29.20 & \best{4.36} & \best{68.95} & \best{0.6483} & \tbest{0.6788} \\
& ResShift-S15 & \best{24.65} & 0.6181 & 0.3349 & 0.2213 & 36.11 & 6.82 & 61.09 & 0.5454 & 0.6071 \\
\cline{2-11}
& SinSR-S1 & \tbest{24.41} & 0.6018 & 0.3240 & 0.2066 & 35.57 & 6.02 & 62.82 & 0.5386 & 0.6471 \\
& OSEDiff-S1 & 23.72 & 0.6108 & \tbest{0.2941} & 0.1976 & 26.32 & 4.71 & 67.97 & 0.6148 & 0.6683 \\
& S3Diff-S1 & 23.52 & 0.5949 & \best{0.2581} & \best{0.1725} & \best{19.66} & 4.74 & \tbest{68.01} & \sbest{0.6318} & \best{0.7012} \\
& \textbf{AdcSR-S1 (Ours)} & 23.74 & 0.6017 & \sbest{0.2853} & \sbest{0.1899} & \tbest{25.52} & \best{4.36} & 68.00 & 0.6090 & 0.6764 \\
\hline \hline
\multirow{13}{*}{DRealSR} & BSRGAN & \best{28.75} & \tbest{0.8031} & \tbest{0.2883} & 0.2142 & 155.63 & 6.52 & 57.14 & 0.4878 & 0.4915 \\
& Real-ESRGAN & \sbest{28.64} & \sbest{0.8053} & \sbest{0.2847} & \best{0.2089} & 147.62 & 6.69 & 54.18 & 0.4907 & 0.4422 \\
& LDL & 28.21 & \best{0.8126} & \best{0.2815} & \tbest{0.2132} & 155.53 & 7.13 & 53.85 & 0.4914 & 0.4310 \\
& FeMASR & 26.90 & 0.7572 & 0.3169 & 0.2235 & 157.78 & \sbest{5.91} & 53.74 & 0.4420 & 0.5464 \\
\cline{2-11}
& StableSR-S200 & 28.03 & 0.7536 & 0.3284 & 0.2269 & 148.98 & 6.52 & 58.51 & 0.5601 & 0.6356 \\
& DiffBIR-S50 & 26.71 & 0.6571 & 0.4557 & 0.2748 & 166.79 & 6.31 & 61.07 & 0.5930 & 0.6395 \\
& SeeSR-S50 & 28.17 & 0.7691 & 0.3189 & 0.2315 & 147.39 & 6.40 & \sbest{64.93} & \tbest{0.6042} & 0.6804 \\
& PASD-S20 & 27.36 & 0.7073 & 0.3760 & 0.2531 & 156.13 & \best{5.55} & \tbest{64.87} & \best{0.6169} & 0.6808 \\
& ResShift-S15 & \tbest{28.46} & 0.7673 & 0.4006 & 0.2656 & 172.26 & 8.12 & 50.60 & 0.4586 & 0.5342 \\
\cline{2-11}
& SinSR-S1 & 28.36 & 0.7515 & 0.3665 & 0.2485 & 170.57 & 6.99 & 55.33 & 0.4884 & 0.6383 \\
& OSEDiff-S1 & 27.92 & 0.7835 & 0.2968 & 0.2165 & \tbest{135.30} & 6.49 & 64.65 & 0.5899 & \tbest{0.6963} \\
& S3Diff-S1 & 27.39 & 0.7469 & 0.3129 & \sbest{0.2108} & \best{119.21} & \tbest{6.17} & 64.16 & \sbest{0.6081} & \best{0.7156} \\
& \textbf{AdcSR-S1 (Ours)} & 28.10 & 0.7726 & 0.3046 & 0.2200 & \sbest{134.05} & 6.45 & \best{66.26} & 0.5927 & \sbest{0.7049} \\
\hline \hline
\multirow{13}{*}{RealSR} & BSRGAN & \best{26.39} & \best{0.7654} & \best{0.2670} & \tbest{0.2121} & 141.28 & 5.66 & 63.21 & 0.5399 & 0.5001 \\
& Real-ESRGAN & 25.69 & \sbest{0.7616} & \sbest{0.2727} & \sbest{0.2063} & 135.18 & 5.83 & 60.18  & 0.5487 & 0.4449 \\
& LDL & 25.28 & \tbest{0.7567} & \tbest{0.2766} & \tbest{0.2121} & 142.71 & 6.00 & 60.82 & 0.5485 & 0.4477 \\
& FeMASR & 25.07 & 0.7358 & 0.2942 & 0.2288 & 141.05 & 5.79 & 58.95 & 0.4865 & 0.5270 \\
\cline{2-11}
& StableSR-S200 & 24.70 & 0.7085 & 0.3018 & 0.2288 & 128.51 & 5.91 & 65.78 & 0.6221 & 0.6178 \\
& DiffBIR-S50 & 24.75 & 0.6567 & 0.3636 & 0.2312 & 128.99 & 5.53 & 64.98 & 0.6246 & 0.6463 \\
& SeeSR-S50 & 25.18 & 0.7216 & 0.3009 & 0.2223 & 125.55 & \tbest{5.41} & \sbest{69.77} & \sbest{0.6442} & 0.6612 \\
& PASD-S20 & 25.21 & 0.6798 & 0.3380 & 0.2260 & 124.29 & \tbest{5.41} & 68.75 & \best{0.6487} & 0.6620 \\
& ResShift-S15 & \sbest{26.31} & 0.7421 & 0.3460 & 0.2498 & 141.71 & 7.26 & 58.43 & 0.5285 & 0.5444 \\
\cline{2-11}
& SinSR-S1 & \tbest{26.28} & 0.7347 & 0.3188 & 0.2353 & 135.93 & 6.29 & 60.80 & 0.5385 & 0.6122 \\
& OSEDiff-S1 & 25.15 & 0.7341 & 0.2921 & 0.2128 & \tbest{123.49} & 5.65 & \tbest{69.09} & 0.6326 & \tbest{0.6693} \\
& S3Diff-S1 & 25.19 & 0.7315 & 0.2707 & \best{0.1994} & \best{110.34} & \best{5.33} & 67.92 & \tbest{0.6398} & \best{0.6761} \\
& \textbf{AdcSR-S1 (Ours)} & 25.47 & 0.7301 & 0.2885 & 0.2129 & \sbest{118.41} & \sbest{5.35} & \best{69.90} & 0.6360 & \sbest{0.6731} \\
\shline
\end{tabular}}
\label{tab:comp_quantitative}
\end{table*}

\subsection{User Study}  
To further evaluate the effectiveness of our AdcSR, we conduct a user study comparing four one-step diffusion-based Real-ISR methods, including SinSR, OSEDiff, S3Diff, and AdcSR. We employ sixteen LR images from the DIV2K-Val dataset, shown in a thumbnail form in Fig.~\ref{fig:user_study_LR_images}. Thirty-two expert researchers are invited to choose the best super-resolution image for each test sample based on two equally weighted criteria: (1) perceptual quality, focusing on clarity, detail, and realism, and (2) content consistency with the LR input, including alignment in image structure and texture.

As reported in Tab.~\ref{tab:user_study_voting_rate}, AdcSR achieves a high voting rate of 31\%, comparable to those of 29\% and 33\% obtained by OSEDiff and S3Diff, both of which employ the complete SD models. Although SinSR has fewer total parameters, its super-resolution quality can be less favorable, as reflected by a lower voting rate of 7\%. These results validate that our compressed diffusion-GAN hybrid maintains highly competitive Real-ISR performance while achieving 4.3$\times$, 3.7$\times$, and 9.3$\times$ faster inference than SinSR, OSEDiff, and S3Diff, respectively, and reducing computation by 81\%, 78\%, and 81\% in GMACs, thus verifying its appealing efficiency.

\begin{table}[!t]
\setlength{\tabcolsep}{2pt}
\centering
\caption{\textbf{Efficiency comparison on a flagship mobile device}, \textbf{\textit{Qualcomm SM8750} (Snapdragon 8 Gen 4)}, for super-resolving an LR input image of size $128\times 128$ with a scaling factor of 4.}
\vspace{-5pt}
\resizebox{\linewidth}{!}{
\begin{tabular}{l|ccc}
\shline
Method & Latency (ms)↓ & Memory (MB)↓ & Storage (MB)↓ \\
\hline \hline
OSEDiff & 1647 & 1777 & 1693 \\
\textbf{AdcSR (Ours)} & \best{65} & \best{510} & \best{435} \\
\hline
Reduction Rate (\%) & 97 & 71 & 74 \\
\shline
\end{tabular}}
\label{tab:efficiency_comp_mobile}
\end{table}

\section{Efficiency Evaluation on Mobile Device}  
We conduct an efficiency comparison of the proposed AdcSR method against its teacher model, OSEDiff, on a flagship mobile platform, \textbf{\textit{Qualcomm SM8750} (Snapdragon 8 Gen 4)} \cite{qsp}, utilizing the Hexagon Digital Signal Processor (DSP). All models are evaluated using the Qualcomm AI Engine Direct Software Development Kit (SDK) \cite{sdk} with 8-bit weights and 16-bit activations (W8A16) quantization for fair comparison. The results reported in Tab.~\ref{tab:efficiency_comp_mobile} demonstrate that AdcSR significantly outperforms OSEDiff in both speed and resource efficiency. Specifically, AdcSR achieves a 25$\times$ acceleration in inference latency, reduces memory footprint by 71\% (from 1.7GB to 0.5GB), and decreases storage requirements by 74\% (from 1.7GB to 0.4GB). These savings are substantial for practical deployment on mobile devices, where resources are typically constrained. To summarize, AdcSR advances beyond previous SD-based one-step Real-ISR models, providing a more efficient, cost-effective solution for real mobile applications.

\small
\bibliographystyle{ieeenat_fullname}
\bibliography{ref}

\end{document}